\documentclass[a4paper,twocolumn,superscriptaddress,nofootinbib,accepted=2020-11-16]{quantumarticle}
\pdfoutput=1

\usepackage{graphicx}% Include figure files
\usepackage{dcolumn}% Align table columns on decimal point
\usepackage{bm}% bold math
\usepackage{physics} %not included in the template initially
\usepackage{bbold}%not included in the template initially
\usepackage{mathtools}%not included in the template initially
\usepackage{color}%not included in the template initially
\usepackage{tikz}
\usepackage{tikz-3dplot}
\usepackage{tikz-cd,comment}
\usetikzlibrary{angles,quotes, arrows.meta}
\usepackage{amsmath,amssymb,amsthm}
\usepackage{hyperref}% add hypertext capabilities
\usepackage[numbers,sort&compress]{natbib}
\def\A{{\sf A}}
\def\B{{\sf B}}
\def\C{{\sf C}}
\def\sR{{\sf R}}

\def\H{\mathcal H}

\def\H{{\mathcal H}}

\def\Z{\mathbb Z}
\def\R{\mathbb R}

\def\sR{{\sf R}}

\def\Diff{{\rm Diff}}

\def\SWAP{{\rm SWAP}}

\def\av{{\rm average}}

\def\S{{\sf S}}
\def\W{{\sf W}}
\def\F{{\sf F}}

\def\0{{\bf 0}}
\def\1{{\bf 1}}
\def\2{{\bf 2}}
\def\3{{\bf 3}}
\def\4{{\bf 4}}

\def\bC{{\mathbb C}}
\def\I{{\mathbb 1}}

\def\SU{\mathrm{SU}}

\def\U{\mathrm{U}}
\def\rU{\mathrm{U}}

\def\SO{\mathrm{SO}}
\def\TRUNC{{\rm T}}
\def\REL{{\rm  R}}

\newtheorem{example}{Example}
\newtheorem{principle}{Principle}
\newtheorem{lemma}{Lemma}

\newtheorem{theorem}{Theorem}

\begin{document}

\title{Quantum reference frames for general symmetry groups}

\author{\mbox{Anne-Catherine de la Hamette}}
\orcid{0000-0003-4811-822X}
\email{\ \ \mbox{annecatherine.delahamette@univie.ac.at}}
\affiliation{%
Institute for Theoretical Physics, ETH Z\"urich, Wolfgang-Pauli-Str. 27, 8093 Z\"urich, Switzerland}
\affiliation{Perimeter Institute for Theoretical Physics, 31 Caroline St. N, Waterloo, Ontario, N2L 2Y5, Canada}%

\author{Thomas D. Galley}
\orcid{0000-0002-8870-3215}
\email{tgalley1@perimeterinstitute.ca}
\affiliation{Perimeter Institute for Theoretical Physics, 31 Caroline St. N, Waterloo, Ontario, N2L 2Y5, Canada}%

\date{23 November 2020}
\begin{abstract}
A fully relational quantum theory necessarily requires an account of changes of quantum reference frames, where quantum reference frames are quantum systems relative to which other systems are described. By introducing a relational formalism which identifies coordinate systems with elements of a symmetry group $G$, we define a general operator for reversibly changing between quantum reference frames associated to a group $G$. This generalises the known operator for translations and boosts to arbitrary finite and locally compact groups, including non-Abelian groups.  We show under which conditions one can uniquely assign coordinate choices to physical systems (to form reference frames) and how to reversibly transform between them, providing transformations between coordinate systems which are `in a superposition' of other coordinate systems. We obtain the change of quantum reference frame from the principles of relational physics and of coherent change of reference frame. We prove a theorem stating that the change of quantum reference frame consistent with these principles is unitary if and only if the reference systems carry the left and right regular representations of $G$. We also define irreversible changes of reference frame for classical and quantum systems in the case where the symmetry group $G$ is a semi-direct product $G = N \rtimes P$ or a direct product $G = N \times P$, providing multiple examples of both reversible and irreversible changes of quantum reference system along the way. Finally, we apply the relational formalism and changes of reference frame developed in this work to the Wigner's friend scenario, finding similar conclusions to those in relational quantum mechanics using an explicit change of reference frame as opposed to indirect reasoning using measurement operators.
\end{abstract}

%\keywords{Suggested keywords}%Use showkeys class option if keyword display desired
\maketitle

\section{Introduction} \label{sec: introduction}
In quantum mechanics, physical systems are implicitly described relative to some set of measurement devices. When writing down the quantum state of a system of interest, say a spin-$1/2$ system in the state $\ket{\uparrow_z}$, we mean that the state of the system is `up' relative to a specified direction $\hat{z}$ in the laboratory. In practice, this direction will be associated to a macroscopic physical system in the lab. If we assume that quantum mechanics is a universal theory and therefore applicable at all scales, the systems we make reference to to describe quantum systems should eventually be treated quantum mechanically as well. Reference systems that are themselves treated as quantum systems are referred to as quantum reference frames. Following the success of Einstein's theory of relativity and its inherently relational nature, one may seek to adopt a relational approach to quantum theory as well. In such an approach, most physically meaningful quantities are relational, i.e. they only take on well defined values once we agree on the reference system (or the observer) relative to which they are described. In his papers \cite{Rovelli_1991, Rovelli1996}, Rovelli suggested that quantum mechanics is a complete theory about the description of physical systems relative to other physical systems. In his Relational Quantum Mechanics (RQM) he rejected the idea of observer-independent states of systems and values of observables. The importance of changes of reference frame in special and general relativity suggests the development of an account of changes of quantum reference frame in RQM. Such an account is given in the present work.

Recently, there has been an increased interest in analysing spatial and temporal quantum reference frames and in establishing a formalism that allows to switch between different perspectives \cite{vanrietvelde2018change, hoehn_how_2018,hoehn_switching_2019, Giacomini_2019, castroruiz2019time}. 
The present work is partially based on these approaches which define changes of quantum reference frames for systems that transform under the translation group (in space and time) and the rotation group in three dimensional space. As opposed to other more standard approaches~\cite{aharonov_quantum_1984,Bartlett_2007,Angelo_2012,loveridge_symmetry_2018}, this formalism stresses the lack of an external reference frame from the outset and defines states of subsystems relative to another subsystem. In standard approaches to quantum reference frames \cite{Bartlett_2007, Angelo_2012}, one often starts from a description relative to an external reference frame and removes any dependence on this reference frame by applying a $G$ twirl (a group averaging over all possible configurations of the external reference frame). In some cases, one can refactor the total Hilbert space into global and relational subsystems and trace out the global degrees of freedom~\cite{Bartlett_2007,Angelo_2012,galilei_pereira_2015,quantum_smith_2016}. The main emphasis of these standard approaches is often to obtain the physically meaningful (or reference frame independent) quantities, in a similar fashion to identifying noise free subsystems in error correction. In the work of~\cite{vanrietvelde2018change, hoehn_how_2018, hoehn_switching_2019, Giacomini_2019, castroruiz2019time, vantrietvelde_switching_2018, hoehn2019trinity} however emphasis is given on the relational nature of the description (always starting from a state that is given from the viewpoint of one of the subsystems) and the main object of study is the relation between different accounts. We make a similar emphasis in the present work. We abstract the formalism of~\cite{Giacomini_2019} and introduce an approach which makes heavy use of the inherently group theoretic nature of quantum reference frames. This allows us to generalise the known results beyond the translation and rotation groups to arbitrary finite and locally compact groups (including non-Abelian groups).

In Section~\ref{sec: relational approach to quantum theory} we outline the relational approach to quantum theory embraced in the present work as well as give a simple example of a change of reference frame for classical bits and an example of a change of quantum reference frame for qubits. In Section~\ref{sec: Reference frames and symmetry groups} we define the notion of a reference frame in terms of reference systems and coordinate systems, as well as give a full account of active and passive transformations as left and right regular group actions. Combining these we define changes of reference frame under a group $G$ for classical systems with configuration space $X \cong G$. In Section~\ref{sec:quantum_change_of_RF} we extend the classical change of reference frame to quantum systems $L^2(G)$ following the principle of coherent change of reference system; and define a general unitary operator which implements this change of reference system. We prove a theorem stating that only systems carrying a regular representation of $G$ can serve as reference frame, subject to the principle of coherent change of reference system. Following this we extend the change of quantum reference frame operator between $L^2(G)$ systems describing systems which do not carry the right regular representation of $G$. In Section~\ref{sec:Irrev_classical} we define irreversible changes of reference frames for groups $G = N \rtimes P$ and $G  = N \times P$ via a truncation procedure. In Section~\ref{sec:Irrev_quantum} we extend this change of reference frame to quantum reference frames using the principle of coherent change of reference system once more. In Section~\ref{sec: Wigner's friend experiment} we apply the tools developed in the preceding sections to the Wigner's friend thought experiment, providing an explicit change of reference frame from Wigner's description to the friend's. We discuss related work in Section~\ref{sec:discussion} and discuss implications of the present work as well as suggestions for future work. In Section~\ref{sec: conclusion} we give some concluding remarks.

\section{Relational approach to quantum theory} \label{sec: relational approach to quantum theory}

In the construction of a relational formalism of quantum mechanics, an essential task is to write quantum states of systems relative to a specified reference system. We introduce the following notation: $\ket{\psi}_\B^\A $ indicates the state of system $\B$ \emph{relative} to system $\A$.
In contrast to the approach of \cite{Giacomini_2019}, we assign a Hilbert space to the system whose perspective is adopted and assign to it the trivial state, corresponding to the identity element of the group. Hence, by convention, system $\A$ is in a default `zero-state' relative to itself. Once we introduce the notion of symmetry groups and how they enter into the formalism, we will see that this default zero-state corresponds to the identity element of the group that describes the transformations of the system. Thus, to be more precise, one can write
\begin{equation}
    \ket{0}_\A^\A \otimes \ket{\psi}_\B^\A. 
\end{equation}
The upper index refers to the system relative to which the state is given while the lower index refers to the system that is being described (similarly to the perspectival approach of~\cite{Bene_perspectival_2001}). This description does not make use of any external abstract reference frame nor does it assume the existence of absolute, observer-dependent values of physical observables. We observe that since system $\A$ can only ever assign itself a single state there are no state self-assignment paradoxes~\cite{dalla_chiara_logical_1977,breuer_impossibility_1995}.

A natural question to address on the relational approach to quantum theory is how to change reference systems. Namely if the state of $\B$ relative to $\A$ is $\ket \psi^\A_{\B} = \ket{0}_\A^\A \otimes \ket{\psi}_\B^\A$, what is the state $\ket \psi^\B_{\A}$ of $\A$ relative to $\B$? This is the problem which will be addressed in the present work.

Before introducing the general framework we will be using, we give two simple examples of changes of reference frame for relational states. The first is classical and the second its quantum generalisation. These should hopefully provide the reader with an intuitive picture of the general mechanisms at play.

\begin{example}[$\Z_2$ change of classical reference frame]
Let us consider the case where systems can be in two states $\uparrow$ or $\downarrow$. Every system considers themselves to be in the state $\uparrow$ (for example an observer free floating in empty space would always consider the up direction to be aligned from their feet to their head). Consider classical systems where the state relative to $\A$ is $\uparrow _\A^\A   \uparrow _\B^\A \downarrow _\C^\A$. Since $\A$ sees $\B$ in the state $\uparrow$ relative to itself, $\B$ also sees $\A$ in the state $\uparrow$ relative to itself. The state relative to $\B$ is $\uparrow _\B^\B   \uparrow _\A^\B  \downarrow_\C^\B$. If the state relative to $\A$ was instead  $ \uparrow _\A^\A  \downarrow _\B^\A \downarrow _\C^\A$ then since $\A$ views $\B$ in the $\downarrow$ state, this implies that $\B$ views things `upside down' relative to $\A$. The change of perspective would give  $\uparrow _\B^\B  \downarrow _\A^\B \uparrow _\C^\B$. 
\end{example} 

In the next example we give a quantum generalisation of the $\Z_2$ change of reference frame. This is a specific instance of the general changes of quantum reference frame defined in this work.

\begin{example}[$\Z_2$ change of quantum reference frame]\label{ex:z2quant}
Let us consider the case with quantum systems $\bC^2$ with basis $\{ \ket \uparrow , \ket \downarrow\}$. Every system considers themselves to be in the state $\ket \uparrow$. By embedding the classical scenario above with the map $\uparrow \ \mapsto \ket \uparrow$ and $\downarrow \ \mapsto \ket \downarrow$  we can reconstruct the classical example: if the state relative to $\A$ is $\ket \psi^\A_{\B\C} = \ket \uparrow _\A^\A  \ket \uparrow _\B^\A \ket \downarrow _\C^\A$ then the state relative to $\B$ is $\ket{\psi}^\B_{\A \C} =  \ket \uparrow _\B^\B  \ket \uparrow _\A^\B \ket \downarrow_\C^\B$. If the state was  $\ket \phi^\A_{\B\C} = \ket \uparrow _\A^\A  \ket \downarrow _\B^\A \ket \downarrow _\C^\A$ then the change of perspective would give  $\ket{\phi}^\B_{\A \C} =  \ket \uparrow _\B^\B  \ket \downarrow _\A^\B \ket \uparrow _\C^\B$. 

Let us move to the quantum case with an initial state $\ket \tau^\A_{\B \C} =\ket \uparrow_\A^\A  \left(\ket \uparrow_\B^\A + \ket \downarrow_\B^\A \right) \ket \downarrow_\C^\A$. What is the state $\ket \tau^\B_{\A \C}$? First let us observe that  $\ket \tau^\A_{\B \C} = \ket \psi^\A_{\B\C} + \ket \phi^\A_{\B\C}$, and let us assume that changes of quantum reference frame are coherent (they observe the superposition principle). Then the state $\ket{\tau}^\B_{\A \C} = \ket \uparrow _\B^\B \left(  \ket \uparrow _\A^\B \ket \downarrow _\C^\B + \ket \downarrow _\A^\B \ket \uparrow _\C^\B \right)$ which is an entangled state of $\A$ and $\C$.
\end{example} 

The above example made use of the two guiding principles of this work: the principle of \emph{relational physics} and the principle of \emph{coherent change of reference system}. These are defined in Section~\ref{sec:n_identical_reversible}.

Whenever we use phrases such as `from the viewpoint of' or `from the perspective of', we simply mean `relative to'. Although these expressions might imply that the state $\ket{\psi}_\B^\A$ indicates how system $\A$ \emph{perceives} system $\B$, we do not make this interpretation here. System $\A$ acts as the observer in this description but we should note that there is nothing special about an observer system. No interpretation is made as to what the system \emph{sees}. Rather a change of reference system $\A \to \B$ is a change of description from one where $\A$ is at the origin to one where  $\B$ is at the origin.

\section{Classical changes of reference frames associated to symmetry groups} \label{sec: Reference frames and symmetry groups}

\subsection{Reference systems, coordinate choices and changes of reference frame}

A coordinate system is a purely mathematical object, and need not in general be associated to a physical system. A reference frame consists of a physical system (known as a reference system), and a choice of coordinates such that the reference system is at the origin in that coordinate system. For full definitions we refer the reader to Appendix~\ref{app:background}. In this section we define changes of reference frames for classical systems where the configuration space is itself a group $G$. In Section~\ref{sec:Irrev_classical} we will consider cases where this is no longer holds.

We begin by a simple example which illustrates changes of reference frames and the use of group elements for relative coordinates.

\begin{example}[Three particles on a line]\label{ex:translation}
Consider three classical particles $\A, \B$ and $\C$ on a line, with state $s = (x_\A, x_\B, x_\C)$ in some Cartesian coordinate system (here we omit the velocities since we are just interested in translations in space). The coordinate system $x'$ such that $x'_A = 0$ is said to be associated to $\A$. In this coordinate system the particles have state $s = (x'_\A = 0, x'_\B =  x_\B - x_\A, x'_\C = x_\C - x_\A)$. We observe here that the relative coordinates (to $\A$) $x'_\A, x'_\B$ and $x'_\C$ uniquely identify the translation which maps system $\A$ to systems $\A$, $\B$ and $\C$. Namely the relative distance $x'_\B =  x_\B - x_\A$ is the distance needed to translate $\A$ to $\B$. The relative coordinates $x'_\A, x'_\B$ and $x'_\C$ correspond to the symmetry group transformations relating $\A$, $\B$ and $\C$ to $\A$. If we label a translation of distance $d$ by $t_L(d)$, where $t_L(d)x = d +x$, we have the state relative to $\A$ as $s^\A = (t_L(0), t_L(x_\B - x_\A),t_L(x_\C - x_\A))$. The state relative to $\B$ is $s^\B =(t_L(x_\A - x_\B),t_L(0), t_L(x_\C - x_\B))$. These two relative states are themselves related by the transformation $s^\B = s^\A - (x_\B - x_\A)$. We define the right regular action of the translation group $T_R(d) = x -d$. The change of reference frame $\A$ to $\B$ is given by the right regular action $T_R(x_\B - x_\A)$ of the group element $x_\B - x_\A$ mapping  $\A$ to $\B$. 
\end{example}

In the above example the configuration space $\R$ and the symmetry group $T = (\R, +)$ acting on it are equivalent as manifolds. This equivalence is essential for the existence of a well defined reversible change of quantum reference frame. In Sections~\ref{sec:Irrev_classical} and~\ref{sec:Irrev_quantum} we study scenarios where this is no longer the case, and the changes of reference frame are irreversible. Since the results in this paper also apply to finite groups we cover a simple example.

\begin{example}[$\Z_2$]
Let us consider systems with configuration space $X = \{\uparrow, \downarrow\}$. The symmetry group $G = \{I,F\} = \Z_2$ consisting of the identity $I(\uparrow) =\ \uparrow$ and the flip $F(\uparrow) =\ \downarrow$ is the symmetry group of $X$. A state of four systems of the form $s = \uparrow, \uparrow, \downarrow, \downarrow$ can be expressed as $s = I(\uparrow),I(\uparrow),F(\uparrow),F(\uparrow)$. By considering $\uparrow$ as the `coordinate system' we have that the state $s$ has coefficients $(I,I,F,F)$. In the `coordinate system' $\downarrow$ the state $s$ would have coefficients $(F,F,I,I)$.
\end{example}

We observe that in the above examples \emph{group elements} of the global symmetry groups serve as \emph{relative coordinates}. In the next subsection we make this link more explicit. We also note the importance of the one to one correspondence between states and coordinate transformations. We observe that there is always some conventionality in changes of coordinates: one considers only translations on $\R$ for instance, and not all diffeomorphisms of $\R$ as relating different coordinates. In Appendix~\ref{app:background} we give explicit definitions of coordinate systems on a manifold $X$, emphasising that a coordinate system is different to a coordinate chart (typically used in general relativity), a distinction made in~\cite{baez_dimensional_2006}. Roughly speaking a \emph{coordinate system} on a manifold $X$ is an isomorphism $f: X \to Y$ (where $Y$ a known mathematical object used to describe $X$), whereas a \emph{coordinate chart} is a map from the mathematical object $Y$ to the physical object $X$ which need not be an isomorphism (for instance multiple charts $\R^n$ are used to describe a curved $n$-dimensional manifold $X$ in general relativity but there is no isomorphism from $X$ to $\R^n$).

\subsection{General treatment of reversible changes of reference frame} \label{sec: General treatment of reversible changes of reference frame}

Let us extract the general features of the above scenario which allow for well defined reference frames and reversible changes of reference frame. Consider a configuration space $X$ (which is typically a set with a manifold structure) and a group $G$ acting on $X$ such that there is a unique transformation $g \in G$ relating any pair of points (the action is transitive and free). This implies $X \cong G$ (as sets/manifolds), and the action of $G$ on $X \cong G$ is the group multiplication on itself: $G \times G \to G$. This space is a \emph{principle homogeneous space for $G$}, sometimes called a \emph{$G$ torsor}.  We assume $G$ locally compact and thus equipped with a left Haar measure, denoted $dg$. Many groups of interest in physics, such as the Poincar\'e group, the symmetric group, $\SU(d)$ and $\SO(d)$ are locally compact (compact and finite groups are instances of locally compact groups). One exception is the diffeomorphism group of some space-time manifold $M$, which in general is not locally compact. We observe that $X \cong G$ follows from the requirement that there exists a unique transformation $g \in G$ relating any two points in $X$. Note that we will later go beyond such perfect reference frames and consider cases for which the configuration space $X$ differs from the group $G$.

We use the following example from~\cite{tao_compactness_2013} to introduce active and passive transformations on $G$ torsors.

\begin{example}[Single observer and system on $X \cong G$]
Consider an observer $0$ at location $x_0$ on $X \cong G$ and an object $1$ at location $x_1$. Then the unique transformation $g$ such that $gx_0 = x_1$ is the relative location of $1$ relative to $0$. 

An active transformation is a transformation on the object 1. A transformation $h$ on the object 1 is given by the left regular action $x_1 \mapsto x_1' = hx_1$. The relative location is now $k$ where $k x_0 = x_1'$. Using $g x_0 = x_1$ and $h x_1 = x_1'$ we find that $k = h g$: $hg x_0 = h x_1 = x_1'$. Therefore an active transformation by $h$ corresponds to the left regular action of $h$ on the relative location $g$: $g \mapsto hg$. 

A passive transformation is a transformation on the observer $x_0 \mapsto h x_0$. This induces a transformation on the relative location of $1$ to $0$ which we now outline. Consider the case where $g x_0 = x_1$ and a passive transformation $h$ on $0$ is applied while $1$ is left unchanged (at position $x_1)$. We have $x_0 \mapsto x_0' = h x_0$. Then the relative location of $1$ to $0$ is $k$ where $k x_0' = x_1$. Substituting in  $g x_0 = x_1$  and $h x_0 = x_0'$ gives $k h x_0 = g x_0$ implying that that $k h = g$, and therefore $k = g h^{-1}$ (where we remember that since $X \cong G$ there is always a unique $g \in G$ mapping a pair of points in $X$). Writing in full  $g h^{-1} h x_0 = g h^{-1} x_0' = x_1$ and so the relative location of $1$ relative to $0$ is now $g h^{-1}$. A passive transformation $h$ on $0$ corresponds to the right regular action of $h$ on the relative location $g$: $g \mapsto gh^{-1}$.
\end{example}

The left regular action and right regular action on $G \cong X$ are defined as follows:
\begin{align}
\phi_L(g, x) & =gx\ ,\\ 
\phi_R(g, x) & = x g^{-1}.
\end{align}
Both are defined using the group multiplication, where $x \in G$.  These two actions naturally commute, and hence $X \cong G$ carries an action of $G \times G$, with one factor typically being understood as the active and the second as the passive transformations~\cite{Kitaev_2004}. Although $\phi_R$ acts `to the right', it is a left group action: $\phi(gh,x) = x (gh)^{-1} = x h^{-1} g^{-1} = \phi(h,x) g^{-1} = \phi(g,\phi(h,x))$. Here we take $\phi_L$ as active and $\phi_R$ as passive.\footnote{Active and passive transformations are typically defined as either left actions on different spaces (states and coordinates) or a left and a right action on the same space (typically coordinates). In this case ($X \cong G$) they can be defined as left actions on the same space.} 

A given state of $n$ systems is $s =(x_0,x_1,...,x_{n-1})$, where we omit the velocities $\dot{x_i}$ since we are defining changes of reference frame for the `translation' group $G$ on $X \cong G$. This can be expressed as:
\begin{align}
s =  \left(g_0^0 x_0 , g^0_1 x_0, g^0_2 x_0,...,g^0_{n-1} x_0\right)\ ,
\end{align}
where $g^i_j$ is the unique $g \in G$ such that $g^i_j x_i = x_j$, and $e = g^i_i$ the identity element. We observe that  $g^j_k g^i_j = g^i_k$ and $(g_i^j)^{-1} = g_j^i$. Then \emph{the state $s^0$ of the $n$ systems relative to system $0$ is:}
\begin{align}
s^0 = \left(g_0^0 , g^0_1 , g^0_2,...,g^0_{n-1}\right). 
\end{align} 
The state relative to the system $i$ is:
\begin{align}
s^i = \left(g^i_0, g^i_1, g^i_2,...,g^i_{n-1}\right).
\end{align} 
We observe that we can also describe the state relative to hypothetical systems (i.e. relative to a point $x \in X$ which is not occupied by a system). For instance in the above consider a point $x_n \in X$ such that $x_i \neq x_n \ \forall i \in \{0,...,n-1\}$. Then we can write:
\begin{align}
s^n = \left(g^n_0, g^n_1, g^n_2,...,g^n_{n-1}\right). 
\end{align} 
As in the examples given above, we see that \emph{a relative state $s^i$ is given by all the symmetry transformations $g^i_j$ from system $i$ to system $j$ for all $j \in \{0,..., n-1\}$}. There is a unique relative state $s^i$, which is such that particle $i$ is in state $e$. 

A change of reference frame from system $0$ to system $i$ is a map $s^0 \to s^i$. Let us extend the left and right regular actions of $G$ on itself to states:
\begin{align}
\phi_L(g, s) & = (g x_0, g x_1,..., g x_{n-1}) \ ,\\ 
\phi_R(g, s) & = ( x_0 g^{-1},  x_1 g^{-1},...,  x_{n-1} g^{-1}) .
\end{align}  

The transformation $s^0 \to s^i$ is given by the \emph{right regular action} of $g^0_i$: 
\begin{align}
\phi_R(g^0_i, s^0) & = \left(  e g_0^i ,  g^0_1 g^i_0, g^0_2 g^i_0,...,g^0_{n-1} g^i_0 \right) \nonumber \\
& =   \left(  g^i_0,  g^i_1, g^i_2, ... , g^i_{n-1} \right) =  s^i. 
\end{align} 
We observe that this transformation cannot be achieved using the left regular action: there is no elements $g \in G$ such that $\phi_L(g,s^0) = s^i$ (unless $G$ is Abelian). The transformation $s^0 \to s^i$ is a \emph{passive transformation}.

\begin{figure}[h]
    \centering
    \begin{tikzcd}[sep=huge]
    1 \arrow[dr, shift right, "g_0^1", swap]
    \arrow[rr, yshift=0.6ex, "g^1_2 = g_2^0 g_0^1", above]
    && 2\arrow[ll, yshift=-0.6ex, " g^2_1= g_1^0 g_0^2", below] \arrow[dl, shift right, "g_0^2", swap]\\
     & 0 \arrow[ur, shift right, "g_2^0 ", swap] \arrow[ul, shift right, "g_1^0", swap]
    \end{tikzcd}
    \caption{Diagram capturing the relational states between three systems. Each system $i$ assigns the relative state along the arrow point from $i$ to $j$ to system $j$ (and the identity to themselves). For instance system $0$ assigns state $e, g^0_1,g^0_2$. By the right regular action of $g^0_1$, we obtain $g^1_0, e, g^0_2g^1_0 = g^1_2$ which is the relative state assigned by system $1$.}
\end{figure}
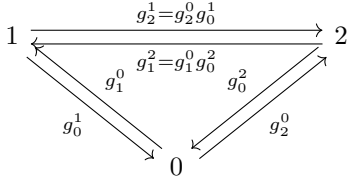

\section{Quantum reference frames associated to symmetry groups}\label{sec:quantum_change_of_RF}

We begin this section by reviewing changes of quantum reference frame for three particles on the line. We then define a quantum change of reference frame operator for $n$ identical systems $L^2(G)$ for arbitrary $G$. This generalises the change of reference frame in~\cite{Giacomini_2019} beyond one parameter subgroups of the Galilean group. 
%We further define a change of reference frame operator for $n-m$ identical $L^2(G)$ systems serving as reference frames and $m$ systems which are not. 
Furthermore we show that it is only the $L^2(G)$ system described so far for which a unitary reversible change of reference system is possible. Finally, we define a change of reference frame operator for $m$ identical $L^2(G)$ systems serving as reference frames and $n-m$ systems which are not. 

\subsection{Comment on finite groups and notation}

All our results apply for finite groups. In this case $L^2(G)$ should be replaced by $\bC[G] \cong \bC^{|G|}$ and integrals $\int_{g \in G} \ketbra{g}{g} dg$ by $\sum_i \ketbra{g_i}{g_i}$. $\bC[G]$ is the vector space freely generated by the elements of $G$, i.e. for which the elements of $G$ form a basis. 

In the cases where it is clear which system is the reference system we sometimes omit the top label for ease of reading. For instance the state $\ket 0_\A^\A \ket{x_1}_\B^\A \ket{x_2}_\C^\A$ is written as $\ket 0_\A \ket{x_1}_\B \ket{x_2}_\C$.

\subsection{The example of $L^2(\R) \otimes L^2(\R) \otimes L^2(\R)$}

Let us first rephrase the known case of the translation group acting on three particles on the line~\cite{Giacomini_2019} in the formalism outlined above.

Take the translation group $T=(\R,+)$ and three systems $\A$, $\B$ and $\C$ whose joint state space is $L^2(\R) \otimes L^2(\R) \otimes L^2(\R)$. Let us for instance consider the state
\begin{equation}
    \ket{0}_\A \ket{x_1}_\B \ket{x_3}_\C\ ,
\end{equation}
which is the state of three perfectly localised systems, described using a coordinate system centred on system $\A$. In standard quantum mechanics, when changing from a classical, highly localised reference frame at the position of $\A$ to another classical reference frame localised at $\B$ translated by an amount $x_1$, one simply applies the translation operator $\hat{T}(-x_1)=e^{ix_1 (\hat{p}_\A +\hat p_\B + \hat p_\C)}$ to the state of the three systems, where $\hat p_\A$ is the momentum operator for system $\A$ and similarly for $\hat p_\B$, $\hat p_\C$ and systems $\B$ and $\C$. This shifts the state to:
\begin{equation}
    \ket{-x_1}_\A \ket{0}_\B \ket{x_3-x_1}_\C.
\end{equation}
In the previous language we have $g^\A_\B = x_1$ and $g^\A_\C = x_3$. The action of $\hat{T}(-x_1)$ corresponds to the right action of  $g^\A_\B = x_1$. 

The next step is to begin with a state of the following form:
\begin{align}\label{eq:example_state_R}
    \ket{0}_\A \frac{1}{\sqrt{2}}(\ket{x_1}+\ket{x_2})_\B \ket{x_3}_\C\ ,
\end{align}
which is described by a coordinate system localised at $\A$. What is the change of reference frame $\A \to \B$ in this case? How can one describe classical coordinates which assign state $\ket 0_\B$, when $\B$ is not localised relative to $\A$? A standard translation of all states will not work. 

Following the reasoning presented in \cite{Giacomini_2019} we assume that the change of perspective obeys the principle of superposition. Namely the state of Equation~\eqref{eq:example_state_R} is an equally weighted superposition of the states $\ket{0}_\A \ket{x_1}_\B \ket{x_3}_\C $ and $\ket{0}_\A \ket{x_2}_\B \ket{x_3}_\C $. The change of reference frame for each of these states individually is obtained by translating by $-x_1$ and $-x_2$ respectively.

Assuming that the superposition principle applies to changes of reference systems, the state described in coordinates `localised' at system $\B$ is just the superposition of the classical states obtained by translation by $-x_1$ and $-x_2$. This leads to the following state of the joint system from the viewpoint of $\B$:
\begin{equation}
    \ket{0}_\B \frac{1}{\sqrt{2}}(\ket{-x_1}_\A \ket{x_3-x_1}_\C+\ket{-x_2}_\A \ket{x_3-x_2}_\C).
\end{equation}
When changing between the viewpoints of \emph{quantum} systems, we apply a weighted translation of the states of systems, dependent on the state of the new reference frame whose viewpoint we are adopting. For the state given above, this means applying a translation for the state of $\B$ being $\ket{x_1}$ and one for it being $\ket{x_2}$.

\begin{figure}[h]
    \centering
    \includegraphics[width=0.9\linewidth]{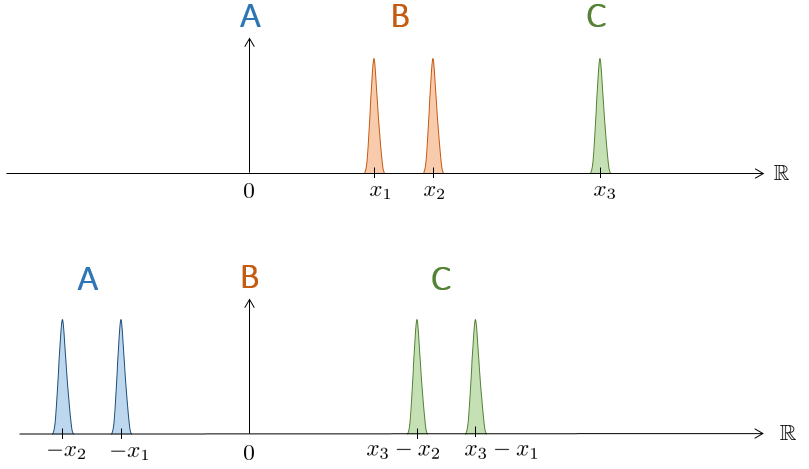}
    \caption{Example for translation group: In the upper subfigure, the state of the three systems is given from the perspective of system $\A$. The lower subfigure shows the state relative to $\B$.}
    \label{fig: translation example}
\end{figure}

The states of $\A$ and $\C$ become entangled relative to $\B$. We see that to perform this change of reference frame, the state of $\A$ is mapped to the inverse of the group element associated with the old state of $\B$. Also for each state of $\B$, the state of $\C$ is shifted respectively. Hence, for the translation group on the real line, the reference frame change operator is
\begin{align} \label{eq: RF change translation group}
  U^{\A \to \B} = &\SWAP_{\A,\B} \circ \nonumber \\
  & \int dx_i dx_j \ketbra{-x_i}{x_i}_\B \otimes \mathbb{1}_\A  \otimes \ketbra{x_j-x_i}{x_j}_\C.
\end{align} 
This operator performs exactly the same reference frame change as the operator given in \cite{Giacomini_2019}:
\begin{equation}
    \hat{S}^{\A \to \B} = \hat{\mathcal{P}}_{AB} e^{i/\hbar \hat{x}_B\hat{p}_C}\ ,
\end{equation}
where $\hat{\mathcal{P}}_{AB}$ is the so-called parity-swap operator. It acts as $\hat{\mathcal{P}}_{AB}\psi_B(x)=\psi_A(-x)$. The proof of this is given in Appendix \ref{app: our operator vs Giacomini operator}.

\subsection{$n$ identical systems $L^2(G)$}~\label{sec:n_identical_reversible}

Consider a configuration space $X \cong G$ and $n$ systems each with associated Hilbert space $\H_i \cong L^2(G)$ for $G$ continuous (or $\bC^{|G|}$ for $G$ finite):
\begin{align}
      G &\to L^2(G) \ ,  \nonumber\\
    g_i &\mapsto \ket{g_i}.
\end{align}
$L^2(G)$ is the space of square integrable functions $G \to \bC$.

The left and right action of $G$ onto itself induces the \emph{left regular} and \emph{right regular} representation of $G$ on each $\H_i$. For a given $\H_i$ this representation acts on the basis $\{\ket g \}$ as:
\begin{align}
U_L(g_2): \ket{g_1} &\mapsto \ket{g_2 g_1} \ ,\\
U_R(g_2): \ket{g_1} &\mapsto \ket{g_1 g_2^{-1}}.
\end{align}
An arbitrary basis state of the $n$ systems is:
\begin{align}
\ket \psi = \ket{g_0}_0 \ket{g_1}_1 ... \ket{g_{n-1}}_{n-1}. 
\end{align}
Following the classical case, the choice of coordinates on $G$ associated to $\H_0$ is:
\begin{align}
\ket \psi^0 = \ket{e}_0 \ket{g^0_1}_1 ... \ket{ g_{n-1}^0}_{n-1}\ , 
\end{align}
where $g^i_j g_i = g_j$. For general $\H_i$ it is:
\begin{align}
\ket \psi^i =  \ket{g^i_0}_0 \ket{g^i_1}_1 ... \ket e_i ...  \ket{ g_{n-1}^i}_{n-1}. 
\end{align}
The change of coordinate system $\ket \psi^0 \to \ket \psi^i$ is given by $U_R(g^0_i)^{\otimes n}$, when considering orthogonal basis states alone.

Let us observe that the left regular representation on the space of wave functions acts as $\psi(x) \mapsto \psi(g^{-1} x)$ and the right regular representation as $\psi(x) \mapsto \psi(xg)$. This follows from $\int_{x \in G} \psi(x)\ket{gx} dx = \int_{x \in G} \psi(g^{-1}x) \ket x dx$ and $\int_{x \in G} \psi(x) \ket{x g^{-1}} dx = \int_{x \in G}\psi(x g) \ket x dx$.  We note that for Lie groups $G$  the objects $\ket g$ are not in $L^2(G)$  and one should typically prefer the representation acting on the wavefunctions.  In the following however we consider the representation acting on the elements $\ket{g}$ in order to describe the continuous and discrete case simultaneously.

%Now following the \emph{fully relational} account embraced in this work we do not assign a global state $\ket \psi \in \H$ and then work out its expression relative to a certain system. 
Unlike some approaches to relational quantum dynamics~\cite{vantrietvelde_switching_2018, vanrietvelde2018change, hoehn2019trinity} we do not assign a global state $\ket \psi \in \H$ and then work out its expression relative to a certain system.
Rather, we begin from a state relative to a system and define changes of reference frame to other systems. We formalise this in the following principle:
\begin{principle}[Relational physics]
Given $n$ systems, states are defined to be relative to one of the systems. A state relative to system $i$ is a description of the other $n-1$ systems, relative to $i$.  
\end{principle}
We observe that this principle does not preclude the existence of a well defined global state of the $n$ systems. 

A superposition state (relative to $0$ in the $G$ product basis) is of the form:
\begin{align}
\ket \psi^0 = & \ket{e}_0  \ket{g_1^0}_1 ... \ket{ g_{n-1}^0}_{n-1} + \ket{e}_0 \ket{h_1^0}_1 ... \ket{ h_{n-1}^0}_{n-1}.
\end{align}
Since the state is in general not a basis state there is no a priori well defined change given by an operator of the form $U_R(g^0_i)^{\otimes n}$. However, following the example of~\cite{Giacomini_2019} one can define a \emph{coherent change of reference frame} operator. We explicitly state this as a principle:
\begin{principle}[Coherent change of reference system]
If $\ket \psi^0 \mapsto \ket \psi^i$ and $\ket \phi^0 \mapsto \ket \phi^i$ then $\alpha \ket \psi^0 + \beta \ket \phi^0 \mapsto \alpha \ket \psi^i + \beta \ket \phi^i$, $\alpha, \beta \in \bC$.
\end{principle}
This implies that $\ket \psi^0$ defined above changes to:
\begin{align}
\ket \psi^i = & \ket{e}_i  \ket{g_0^i}_0 \ket{g_1^0 g_0^i}_1 ...  \ket{ g_{n-1}^0 g_0^i}_{n-1} \nonumber \\
& + \ket{e}_i \ket{h_0^i}_0 \ket{h_1^0 h_0^i}_1 ... \ket{ h_{n-1}^0 h_0^i}_{n-1}.
\end{align}
The operator which implements the coherent change of reference systems $0 \to i$ is:
\begin{align}\label{eq:change_of_RF}
U^{0 \to i} = &\SWAP_{0,i} \circ \nonumber \\ & \int_{g^0_i \in G} \ketbra{g^i_0}{g_i^0}_i \otimes \I_0 \otimes U_R(g^0_i)^{\otimes n - 2} dg_i^0.
\end{align}

The following lemmas are proven in Appendix~\ref{app:proofs_of_lems}.

\begin{lemma}\label{lem:unit}
$U^{0 \to i}$ is unitary. 
\end{lemma}

\begin{lemma}\label{lem:inverse}
$\left(U^{0 \to i}\right)^\dagger = U^{i \to 0}$
\end{lemma}

\begin{lemma}\label{lem:transitive}
$U^{i \to j} U^{k \to i} = U^{k \to j}$
\end{lemma}

\subsubsection{Change of reference frame for observables}

The change of reference frame operator also allows us to transform between observables. Namely if system $0$ describes an observable of systems $1,..., n-1$ as $Z^0_{0,1,...,n-1} = \I_0 \otimes Z^0_{1,...,n-1}$ then system $i$ describes the observable as $Z^i_{0,1,...,n-1} = U^{0 \to i} Z^0_{0,1,...,n-1} U^{i \to 0}$.

\subsubsection{$L^2(\U(1)) \otimes L^2(\U(1)) \otimes L^2(\U(1))$}
To illustrate the changes of reference frame described previously we will give an example. Let us consider the symmetry group $\U(1)$ and three particles $\A, \B, \C$ on a circle with associated Hilbert space  $L^2(\U(1)) \otimes L^2(\U(1)) \otimes L^2(\U(1))$. 
In this case, the map from group elements to elements of the Hilbert space is
\begin{align}
 \U(1) &\to L^2(\U(1)) \ , \nonumber \\
 \theta_i &\mapsto \ket{\theta_i} \ ,
\end{align}
with $\theta_i \in [0,2\pi[$ and $\braket{\theta_i}{\theta_j}=\delta(\theta_i - \theta_j)$. 
\begin{figure}[ht]
    \centering
    \includegraphics[width=0.6\linewidth]{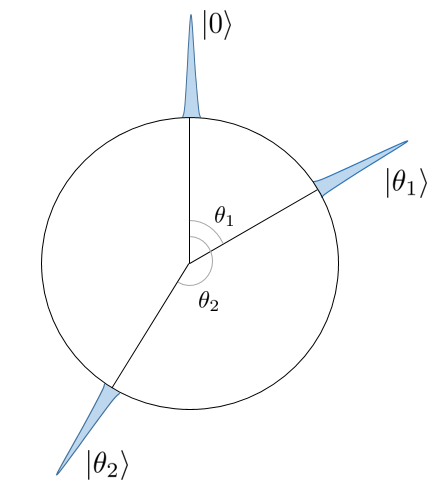}
    \caption{Basis states $\ket{0}, \ket{\theta_1}$ and $\ket{\theta_2}$ of the state space for $L^2(U(1))$.}
    \label{fig: U(1) basis}
\end{figure}
The states $\{ \ket{\theta_i} \vert\ \theta_i \in [0,2\pi[ \}$ are states at all angular positions of the unit circle and form a basis of the Hilbert space $L^2(\U(1))$. A system consisting of three particles on a circle could for instance be in the product state $\ket{0}_\A\ket{\frac{\pi}{2}}_\B\ket{\pi}_\C$ relative to $\A$.

An arbitrary state of the joint system relative to particle $\A$ is given by
\begin{equation}
    \ket{0}_\A \otimes \int d\theta_i d\theta_j\ \psi(\theta_i,\theta_j) \ket{\theta_i}_\B \otimes \ket{\theta_j}_\C \ ,
\end{equation}
where $\ket{e}=\ket{0}$ is the state associated to the identity element of $\U(1)$. As a specific example, take the state
\begin{equation}
    \ket{0}_\A \otimes \left(\sqrt{\frac{1}{3}} \ket{\theta_1} + \sqrt{\frac{2}{3}} \ket{\theta_2}\right)_\B \otimes \ket{\theta_3}_\C.
\end{equation}
Relative to particle $\B$, the state assigned to the joint system would be
\begin{equation}
    \ket{0}_\B  \left(\sqrt{\frac{1}{3}} \ket{-\theta_1}_\A \otimes \ket{\theta_3-\theta_1}_\C + \sqrt{\frac{2}{3}} \ket{-\theta_2}_\A \otimes \ket{\theta_3-\theta_2}_\C\right).
\end{equation}

We see that the state of $\B$ is mapped to the state corresponding to the inverse group element assigned to the old state of $\B$ and the state of $\C$ is shifted respectively. In the end, the labels of $\A$ and $\B $ are swapped. The operator that performs this reference frame change is 
\begin{align}
  U^{\A \to \B} = &\SWAP_{\A,\B} \circ \nonumber \\
  &  \int d\theta d\theta' \ketbra{-\theta}{\theta}_\B \otimes \mathbb{1}_\A \otimes \ketbra{\theta'-\theta}{\theta'}_\C \nonumber\\
  = & \SWAP_{\A,\B} \circ  \int d\theta d\theta' \ketbra{-\theta}{\theta}_\B \otimes \mathbb{1}_\A \otimes U_R(\theta)_\C.
\end{align}

\begin{figure}[h]
    \centering
    \includegraphics[width=0.95\linewidth]{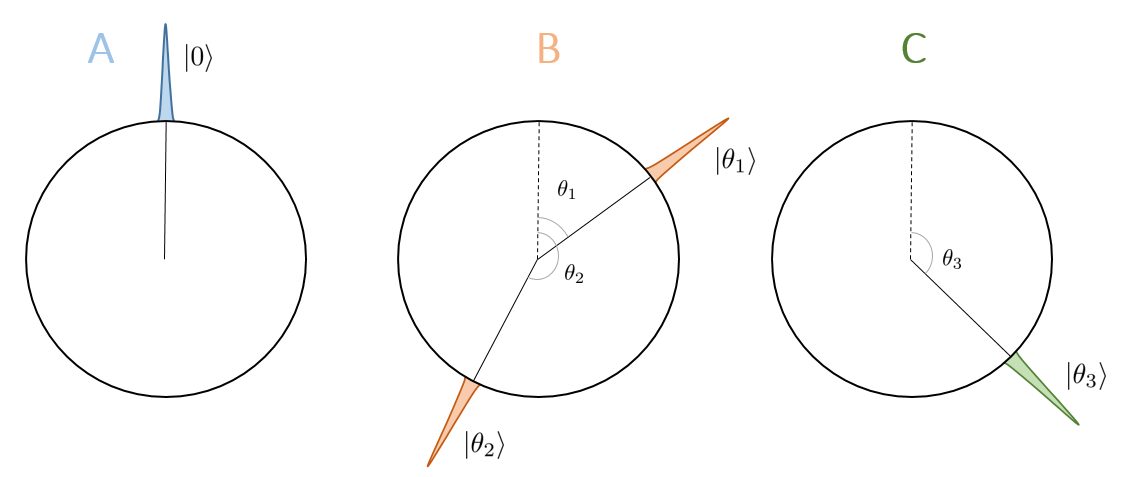}
    \caption{Example of three particles on a circle.}
    \label{fig: tree particles on a circle}
\end{figure}

\subsection{Unitarity of change of reference frame operator}

The change of reference frame defined previously is highly constrained: it applies only to systems $L^2(G)$ with symmetry group $G$. One could ask whether one could define similar changes of reference frame for systems $\H \not \cong L^2(G)$ with states $\ket{\psi(g)}$ carrying two representations: $U_L(h) \ket{\psi(g)} = \ket{\psi(hg)}$ and $U_R(h) \ket{\psi(g)} = \ket{\psi(gh^{-1})}$. 

First consider the symmetry group $\U(1)$. Our results show that for a change of reference frame to recreate our classical intuitions one needs systems $L^2(\U(1))$ which carry the right regular representation. However one may wonder whether one could use qubits with states along the $X-Y$ plane $\ket \theta = \cos(\theta/2) \ket 0 + \sin(\theta/2) \ket 1$ as reference systems which transform in a manner which obeys the classical change of reference frame.

We first provide an example to show that this breaks linearity of the change of reference frame operator for the case of rebits before proving a general result.

\begin{example}[Three qubits with $\U(1)$ group action]
Consider three qubits with states restricted to real valued superpositions: $\ket \theta = \cos(\theta/2) \ket 0 + \sin(\theta/2) \ket 1$ (sometimes known as rebits). The space of pure states of the three systems is  $\U(1) \otimes \U(1) \otimes \U(1)$. We apply our classical intuition of how a reference frame change should act. Let the initial state of the three systems relative to $\A$ be
\begin{equation}
  \ket{\psi(0)}_\A \otimes \ket{\psi(\theta)}_\B \otimes \ket{\psi(\theta')}_\C.
\end{equation}
The map $\psi$ takes the group element $\theta$ of $\U(1)$ to the state in the two-dimensional Hilbert space:
\begin{align}
 \psi:\  \U(1) &\to \mathcal{H} \nonumber \\
    \theta &\mapsto \cos(\theta/2) \ket{0} + \sin(\theta/2) \ket{1}.
\end{align}
We want this state to be mapped to the final state relative to $\B$:
\begin{equation}
    \ket{\psi(-\theta)}_\A \otimes
  \ket{\psi(0)}_\B \otimes \ket{\psi(\theta'-\theta)}_\C.
\end{equation}
This corresponds to our intuition of what should happen when one changes from the viewpoint of $\A$ to the viewpoint of $\B$. Writing out this transformation in the rebit basis corresponds to the map:
\begin{align}\label{eq: three rebits map}
& \ket{0}_\A \otimes (\cos(\theta/2) \ket{0} + \sin(\theta/2) \ket{1})_\B \nonumber\\
&\otimes  (\cos(\theta'/2) \ket{0} + \sin(\theta'/2) \ket{1})_\C \nonumber \\
 \mapsto & (\cos(-\theta/2) \ket{0} + \sin(-\theta/2) \ket{1})_\A \otimes \ket{0}_\B \nonumber \\
& \otimes (\cos((\theta'-\theta)/2) \ket{0} + \sin((\theta'-\theta)/2) \ket{1})_\C.
\end{align}

On the other hand, considering the basis states of the joint Hilbert spaces and assuming the map is linear, the following should hold:
\begin{eqnarray}
 \ket{000}_{\A\B\C} \mapsto &\ket{000}_{\A\B\C} &\ \theta = \theta' = 0 \ , \nonumber \\
 \ket{001}_{\A\B\C} \mapsto &\ket{001}_{\A\B\C} &\ \theta = 0,\  \theta' = \pi\ , \nonumber \\
 \ket{010}_{\A\B\C} \mapsto &\ket{101}_{\A\B\C} &\ \theta = \pi,\  \theta' = 0\ , \nonumber \\
 \ket{011}_{\A\B\C} \mapsto &-\ket{100}_{\A\B\C} &\ \theta = \theta' = \pi.
\end{eqnarray}

When comparing the coefficients in the map \eqref{eq: three rebits map}, one sees that the reference frame change cannot be linear. This means that the operator describing the change from one rebit reference system to another one is non-linear. As this non-linearity causes issues concerning the invariance of probabilities under reference frame change we conclude that rebits cannot serve as reference frames that allow to reversibly transform between each other.
\end{example}

Given $n$ classical systems with configuration space $X \cong G$ acted on by a symmetry group $G$ we have shown how to define states relative to these systems, and to transform between them using the left and right regular action of $G$ on $X$. 

The case $L^2(G)$ is a very specific `encoding' of $G$ into a quantum system. It is a natural choice, in that the classical states are embedded into orthogonal states of the quantum system.  However one could have an injection $G \to \H$, with $g \mapsto \ket{\psi(g)}$ such that the states $\ket{\psi(g)}$ are not all mutually orthogonal and ask whether a change of reference system can be defined. We require $\H$ to carry two unitary representations  $U_L$ and $U_R$, corresponding to active and passive transformations, such that $U_L(h) \ket{\psi(g)} = \ket{\psi(hg)}$ and $U_R(h) \ket{\psi(g)} = \ket{\psi(gh^{-1})}$, where $\ket{\psi(g)} = \ket{\psi(h)} \leftrightarrow g=h$. Although one would naturally desire them to commute (since active and passive transformations as usually defined act on different spaces and therefore trivially commute), we do not impose this here. The following theorem tells us that the change of reference frame which acts as expected on product states $\ket{\psi(e)}_i \ket{\psi(g_0^i)}_0... \ket{\psi(g_j^i)}_j ...\ket{\psi(g_{n-1}^i)}_{n-1}$  and obeys the principle of coherent change of reference frame is unitary exactly if the states $\ket{\psi(g)}$ it acts on form an orthonormal basis of the Hilbert space. 

\begin{theorem}
Take $n$ identical systems with associated Hilbert spaces $\H_i$ each carrying two representations of $G$: $U_L$ and $U_R$ such that $U_L(h) \ket{\psi(g)} = \ket{\psi(hg)}$ and $U_R(h) \ket{\psi(g)} = \ket{\psi(gh^{-1})}$, where $\ket{\psi(g)} = \ket{\psi(h)} \leftrightarrow g=h$. Then any operator $U$ which performs the change $\ket{\psi(e)}_i \ket{\psi(g_0^i)}_0... \ket{\psi(g_j^i)}_j ...\ket{\psi(g_{n-1}^i)}_{n-1} \mapsto \ket{\psi(e)}_j \ket{\psi(g_0^j)}_0... \ket{\psi(g_i^j)}_i ... \ket{\psi(g_{n-1}^j)}_{n-1}$ and obeys the principle of coherent change of reference system is unitary if and only if the representations $U_L$ and $U_R$ are the left and right regular representations acting on states $\ket{\psi(g)}$ which form an orthonormal basis of $\H_i$ (or a subspace thereof).
\end{theorem}

The proof can be found in Appendix \ref{app: unitarity of RF change operator}.

\subsection{$m$ $L^2(G)$ systems describing $n-m$ systems}

Let us consider the case where reference systems $L^2(G)$ describe systems of a different type. The total Hilbert space is $L^2(G)^{\otimes m} \otimes \H^{\otimes n-m}$ where for simplicity we have assumed the $n-m$ systems to be of the same type (but not $L^2(G)$). The systems $\H$ are such that there exists an injection $\phi$:
\begin{align}
\phi  : &\ G \to \H \nonumber \ , \\
  &\ g \mapsto \ket{\psi(g)} \ ,
\end{align}
and two representations $V_L$ and $V_R$ such that:
\begin{align}
V_L(g) \ket{\psi(h)} & = \ket{\psi(gh)} \ , \\
V_R(g) \ket{\psi(h)} & = \ket{\psi(hg^{-1})}.
\end{align}
To change from reference system $0$ to $i$, where both systems are assumed to be of the type $L^2(G)$, we apply the operator:
\begin{align}\label{eq:change_of_RF m perfect systems}
& U^{0 \to i} = \SWAP_{0,i} \circ \nonumber \\
& \int_{g^0_i \in G} \ketbra{g^i_0}{g_i^0}_i \otimes \I_0 \otimes U_R(g^0_i)^{\otimes m - 2} \otimes V_R(g^0_i)^{\otimes n - m} dg_i^0,
\end{align}
where $U_R$ is the right regular representation acting on the first $m$ $L^2(G)$ systems. 

We observe that not all systems $\H$ which carry a representation of $G$ will be such that there exists an injective map $\phi: g \mapsto \ket{\psi(g)}$. For instance the qubit carries a representation of $\SU(2)$ but there is no injection of $\phi: \SU(2) \to {\rm P}\bC^2$ (where here we emphasise that the pure states of a $\bC^2$ system form ${\rm P}\bC^2$ the projective space of rays). Observe that for $\rU(1)$ there is an injection $\phi: \rU(1) \to {\rm P}\bC^2$. We explore the example of two $L^2(\rU(1))$ systems describing a system $\bC^2$ carrying a representation of $\rU(1)$.

\subsubsection{$L^2(\U(1)) \otimes L^2(\U(1)) \otimes \bC^2$}

Let us adapt the previous example of three particles on a circle to the case in which the third system is a qubit $\H_\C \cong \bC^2$ giving a total Hilbert space of the joint system $L^2(\U(1)) \otimes L^2(\U(1)) \otimes \bC^2$. $\H_\C$  carries a representation of $\rU(1)$ and an injection $\phi: \rU(1) \to \bC^2$. The representation $V_R$ is given by:

\begin{align} \label{eq: unitary rep of the rebit}
    V_R(\theta) = \begin{pmatrix}
\cos(-\theta/2) & -\sin(-\theta/2) \\
\sin(-\theta/2) & \cos(-\theta/2)
\end{pmatrix}
\end{align}

in the $\{\ket{0},\ket{1}\}$ basis and acts by matrix multiplication from the left.  The injection is a map 
\begin{align}
    \phi\ : &\  \U(1) \to \bC^2 \nonumber \ ,\\
     &\  \theta \mapsto  \cos(\theta/2) \ket 0 +  \sin(\theta/2) \ket 1 .
\end{align}

The operator that maps the state relative to $\A$ to the state relative to $\B$ is  
\begin{align}
    U^{\A \to \B} = \SWAP_{\A,\B} \circ \int d\theta \ketbra{-\theta}{\theta}_\B \otimes \mathbb{1}_\A \otimes V_R(\theta)_\C.
\end{align}

As a specific example, consider the state
\begin{equation}
    \ket{0}_\A \ket{\pi}_\B \ket{\psi(\frac{\pi}{2})}_\C
\end{equation}
relative to system $\A$, where $\ket{\psi(\frac{\pi}{2})}=\frac{1}{\sqrt{2}} (\ket{0}+\ket{1})$. From the viewpoint of system $\B$, the state is 
\begin{equation}
   \ket{0}_\B \ket{\pi}_\A \ket{\psi(-\frac{\pi}{2}))}_\C  = \ket{0}_\B \ket{\pi}_\A \frac{1}{\sqrt{2}} (\ket{0}-\ket{1})_\C.
\end{equation}

\subsection{Changes of reference frame for arbitrary identical systems}

The above treatment shows that for any group $G$ one can define a change of quantum reference frame between $n$ identical systems. However given $n$ identical systems can one always find a group allowing for a reversible change of quantum reference frame? Namely for a configuration space $X$, can one always find a binary operation turning it into a group $G \cong X$? In the case of finite systems $\bC^d$ one can pick an orthonormal basis $\ket x$, $x \in \{0,..., d-1\}$ and choose the cyclic group $\Z_d$ acting on $\{0,..., d-1\}$. In the case where $X$ is a countable set one has the group $\Z$. In the case where $X$ is uncountable, the existence of a group $G$ such that $G\cong X$ is equivalent to the axiom of choice~\cite{105440}. We observe that if $X$ has some additional structure (such as being a manifold), then one may not be able to find a group which is isomorphic as a manifold.

\section{Irreversible changes of classical reference frame}\label{sec:Irrev_classical}

In some cases one may not have access to a reference system which can distinguish all elements of the symmetry group. Consider once more the case of particles on $\R$ acted on by the translation group. Given a ruler with a set of marks corresponding only to the subset of integers (i.e. with configuration space $X \cong \Z$), one would not be able to distinguish all possible different configurations of the particles and by extension all possible translations. Such imperfect reference frames, with configuration space $X$ which is a coarse-graining of the group $G$, will lead to irreversible changes of reference frame as we will see in this section.

For a given configuration space $X$ all changes of coordinates are related by a transformation $g \in G$. In the case $X \cong G$ there is a one to one correspondence between points in $X$ and coordinate systems. As such one can identify coordinate systems as systems with configuration space $X$. Namely if system $i$ is in state $x_i \in X$ then one assigns it the unique coordinate system $x_i'$ which maps $x_i \mapsto 0$. However in situations such as the one described previously one has a symmetry group $G$ which is larger than $X$ and there is no unique element in $G$ mapping a point $x_i$ to $0$.  We consider an explicit example of this in the following before describing the general case.

\subsection{$E^+(3) \cong \R^3 \rtimes \SO(3)$}

Let us consider $n$ particles in $\R^3$, with each particle $i$ having state $(x_i,y_i,z_i)$ expressed in Cartesian coordinates $(x,y,z)$.  The set of Cartesian coordinates is acted on by the Euclidean group $E^+(3) = \R^3 \rtimes \SO(3)$. A choice of coordinates $(x',y',z')$ such that $(x_i',y_i',z_i') = (0,0,0)$ is said to be associated to particle $i$ if and only if it is the unique set of coordinates with this property. There are infinitely many such coordinate choices (for instance all coordinate systems which are rotated relative to $(x',y',z')$ will also assign state $(0,0,0)$ to particle $i$). In this case there is no obvious unique manner of associating a coordinate system to a particle.

All Cartesian coordinate systems for $\R^3$ are related by an element $g \in E^+(3)$ where $E^+(3)$ is the Euclidean group. The action of $E^+(3)$ on the set of Cartesian coordinates is transitive and free. To put it visually every element in $E^+(3)$ can be considered as a translation followed by a rotation. Every choice of Cartesian coordinates is associated to a set of orthogonal axes located at some point $r \in \R^3$ with a given orientation. These are all related to the Cartesian coordinates at $(0,0,0)$ in a given orientation by a rotation followed by a translation. We cannot assign a unique coordinate system to each point in $\R^3$. For instance take a coordinate system centred at the origin; then any other coordinate system obtained by rotation about the origin will also assign the state $(0,0,0)$ to the origin. See Figure~\ref{fig: rotated coordinate systems, same origin}.

\begin{figure}
    \centering
\includegraphics[width=0.5\linewidth]{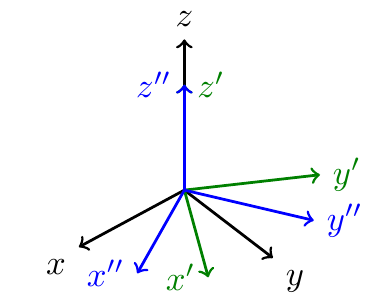}
\caption{Reference frames associated to the same point in $\R^3$.}
\label{fig: rotated coordinate systems, same origin}
\end{figure}

\subsubsection{Enlarging the space of states of the reference systems}

There are multiple ways of addressing this issue. One can say that systems with configuration spaces $\R^3$ (i.e. particles) are not good reference systems for $E^+(3)$. Rather one should choose systems with a larger configuration space. This is what is typically done, where we choose solid bodies in $\R^3$ as reference systems. Since solid bodies have an orientation (unlike points), which is to say that rotating a solid body changes its state, they have configuration space $E^+(3)$. One can assign a unique coordinate system to every state $x \in X \cong E^+(3)$ of a solid body. An example of a solid body would be three physical orthogonal axes in $\R^3$ labelled $1$, $2$ and $3$. For a given state $x$ of these three physical axes one can associate the coordinate system which assigns $+x, +y$ and $+z$ to the axes $1$, $2$ and $3$. Using this approach would allow us to make use of the results of the previous section. 

\begin{figure}
    \centering
    \includegraphics[width=0.35\textwidth]{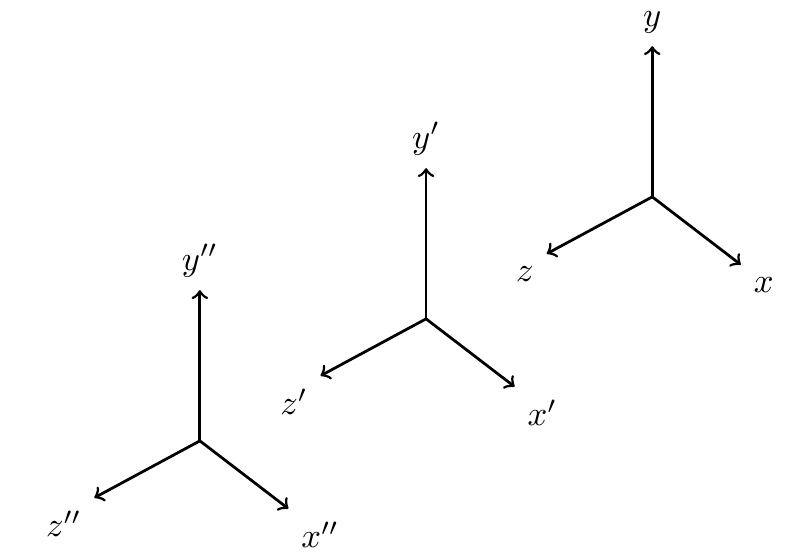}
\caption{For each point in $\R^3$ a representative member of all reference frames centred at the point is chosen. Here the representative member is chosen so that each representative member has the same orientation. This ensures that the closure of the set of transformations relating the different reference frames is $T(\R^3)$ and not a larger group.}
\label{fig: ref same orient}
\end{figure}

However one could also keep the reference systems as having configuration space $X$ but rather assign to each state $x$ the equivalence class of coordinate systems centred on $x$. One can either choose a \emph{representative member} of the equivalence class (in the above case one can fix all coordinate systems to have a given orientation as in Figure~\ref{fig: ref same orient}) or one could average over the possible elements of the equivalence class. 

\subsubsection{Representative element of each equivalence class}

In the case where all the reference systems have configuration space $\R^3$, it makes sense to assign to each point $x \in \R^3$ a unique coordinate system centred on that point (from the equivalence class of coordinate systems centred on that point). 

Take $K$ to be the subset of transformations which relates these coordinate systems. We require that $K$ is a group in order for us to have a well defined change of coordinate system. If $K$ is not a group, then by composing different elements in $K$ we can obtain a group $G'$ (which is larger than $K$ as a set). This symmetry group will take coordinate systems we have selected to coordinate systems which we have not chosen. 

In order for all representative members (i.e. the coordinate system we chose to be associated to each point) to be related by a group $K \subset E^+(3)$ and for the representative members to be closed under the action of $K$ one can choose them to all have the same orientation (i.e. be related by just translations). In this case the symmetry group relating coordinate choices becomes $K \cong \R^3$ once more.

\subsection{$G =  N \rtimes P$ and $G = N \times P$: truncation}

Let us consider the case where there are systems with configuration space $G$ and systems with configuration space $N$, where $G = N \rtimes P$ or $G = N \times P$. In both cases $N$ is normal, and for every $g \in G$ there is a unique $n \in N$ and $p \in P$ such that $g = np$. 

The configuration space $N$ is embedded in $G$ via an embedding map $E: N \to G$, $E: n \mapsto n p_C$ for some constant $p_C$, where the choice of $p_C$ is conventional and is typically chosen to be the identity. For a choice $p_C$, the points $n p_C$ $\forall n \in N$ are related by transformations $n \in N$ (acting to the left). As such the symmetry group of $E(N)$ is $K \cong N$. If the map did not fix a unique convention (for instance $np \mapsto n p_0(n)$) where the image depends on which equivalence class is chosen, then the set $K$ of transformations between the images $E(n)$ would typically not be a group, and its closure would not be isomorphic to $N$ (in some cases it would be the full group $G$).

Take $k$ systems $G$ and $l-k$ systems $N$.  A general state of the $l$ systems is:
\begin{align}
s = \left((g_0, g_1 , ... , g_{k-1}), (g_{k}, ..., g_{l-1}) \right) \ ,
\end{align}
where $g_i \in G$ for $i \in \{0,\ldots, k-1\}$ and $g_j  \in N$ for $j \in \{k,\ldots, l-1\}$. Moreover there is a unique $n_i \in N$ and $p_i \in P$ such that $g_i = n_i p_i$. Here $p_j =e$ for systems $j \in \{k,\ldots, l-1\}$. The description relative to the first $k$ systems and the transformations between them is just the case described in Section~\ref{sec: Reference frames and symmetry groups}. In the following we describe how to change reference system from a system with configuration space $G$ to a system with configuration space $N$.

The embedding of $N \subset G$ is given by $n \mapsto n e$. We define the truncation map:
\begin{align}
\TRUNC:\ &G \to N \ ,\nonumber\\
&g = np \mapsto n  \ ,
\end{align}
and the map $\REL_G^i$:
\begin{align}
\REL_G^i :(g_0,..., g_{l-1}) \mapsto (g^i_0,..., g^i_{l-1}) \ .
\end{align}
Then, the relative state $s^0 = \REL_G^0(s)$ is:
\begin{align}
s^0 & =  \left((e, g_1^0 , ... , g_{k-1}^0), (g_{k}^0, ..., g_{l-1}^0) \right) \nonumber\\
& = \left((e, n_1^0 p_1^0 , ... , n_{k-1}^0 p_{k-1}^0), (n_{k}^0 p_{k}^0, ..., n_{l-1}^0 p_{l-1}^0) \right),
\end{align}
where $n_i^j p_i^j = g_i^j$. 

%Observe that for $j \in \{m,\ldots, n-1\}$ we have $g_j^0 g_0 = n_j$ which implies $g_j^0  n_0 p_0 = n_j \in N$. }

Let us consider the case of $j \in \{k,\ldots, l-1\}$, i.e. where $g_j = n_j$. Then we have that $g_j^0 = g_j g_0^{-1} = n_j g_0^{-1}$. Now let us observe that for all $g$ there is a unique decomposition into $g = np$, and so we write  $g_j^0 = n_j^0 p_j^0$. 
We now work out this $n_j^0 p_j^0$ in terms of $n_j$ and $g_0 = n_0 p_0$ using the two following equalities:
\begin{align*}
g_j^0 & = n_j g_0^{-1} = n_j p_0^{-1} n_0^{-1} \ ,\\
g_j^0 & = n_j^0 p_j^0 \ .
\end{align*}
Combining these gives: 
\begin{align*}
n_j p_0^{-1} n_0^{-1}  = n_j^0 p_j^0.
\end{align*}
Now let us introduce an identity $p_0 p_0^{-1}$ on the RHS: 
\begin{align*}
n_j p_0^{-1} n_0^{-1} p_0 p_0^{-1} = n_j^0 p_j^0\ ,
\end{align*}
and  observe that  $g n g^{-1} \in N$ for all $g \in G$, which implies that $p_0^{-1} n_0^{-1} p_0 \in N$, in turn implying that $n_j p_0^{-1} n_0^{-1} p_0 \in N$. Since the decomposition of $g_j^0$ into $n_j^0 p_j^0$ is unique, this implies that $n_j^0 = n_j p_0^{-1} n_0^{-1} p_0 $ and $p_j^0 = p_0^{-1}$ for all $j \in \{k,\ldots, l-1\}$.
Therefore
\begin{align}
s^0 & =  \left((e, g_1^0 , ... , g_{k-1}^0), (g_{k}^0, ..., g_{l-1}^0) \right)\nonumber\\
& = \left((e, n_1^0 p_1^0 , ... , n_{k-1}^0 p_{k-1}^0), (n_{k}^0 p_C, ..., n_{l-1}^0 p_C) \right),
\end{align}
where $p_C = p_0^{-1}$.

For particle $j$ with configuration space $N$, the state $s^j$ is:
\begin{align}
s^j =\left((n^j_0, n_1^j , ... , n_{k-1}^j ), (n_{k}^j , ..., n_{l-1}^j ) \right),
\end{align}
which is obtained from $s^0$ by the map $(\Gamma_R(n^0_j) \circ \TRUNC)$. Here $\Gamma_R(n)$ is just shorthand for the right regular action: $\Gamma_R(n) (g_0,...,g_{l-1}) = \phi_R(n,(g_0,...,g_{l-1})) = (g_0n^{-1},...,g_{l-1} n^{-1})$. We have $n_i^0 n^j_0 = n^j_i$.

We observe that all states $s^0$ of the form $\left((n_0^0 p_0^0, n_1^0 p_1^0 , ... , n_{k-1}^0 p_{k-1}^0), (n_{k}^0p_{k}^0, ..., n_{l-1}^0p_{l-1}^0) \right)$ for all $p_i^0 \in P$ give the same $s^j$. The change of reference frame $s^0 \mapsto s^j$ is an irreversible change of reference frame.

\subsubsection{$\R \cong \Z \rtimes \rU(1)$}\label{sec:mod_enc}

An example of such a truncation is the `modular truncation' of the translation group: $\R \cong \Z \rtimes \rm{U}(1)$. Instead of distinguishing all position states on the real line, we consider a reference frame that essentially consists of a classical ruler. All points that lie within an interval of length $L$ are mapped to the same point on the ruler (for simplicity, one can choose $L=1$). Hence, the map identifies a subset of elements of $G=\R$ with the same element of $N=\Z$:
\begin{align}
    T:\   \R &\to \Z \nonumber \ ,\\
    x &\mapsto n  \ ,
\end{align}
where $x = nL + p, \ p \in [0,L[, \ n = \left \lfloor{\frac{x}{L}}\right \rfloor \in \Z$. Physically, this truncation can be viewed as coarse-graining; in the sense of resolution, this means we transform from a finer to a coarser resolution.

\begin{figure} [ht]
    \centering
\includegraphics[width=0.45\textwidth]{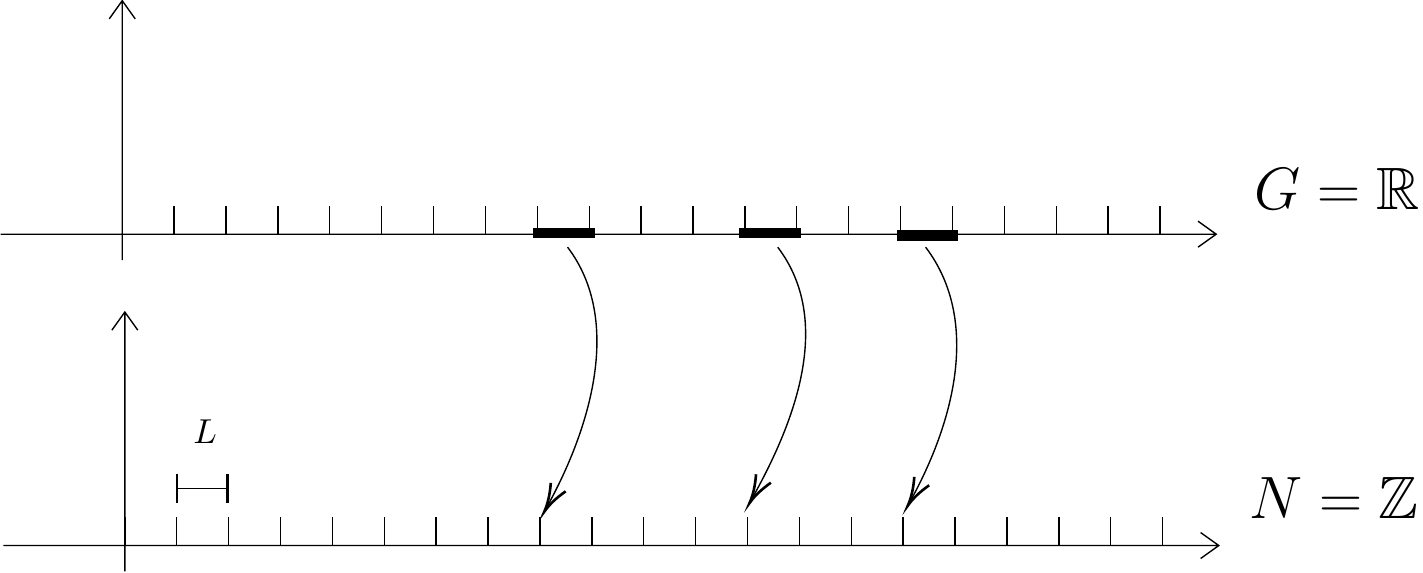}
    \caption{Modular encoding of the real line.}
    \label{fig:modular encoding}
\end{figure}

Consider now three particles on the real line at positions $x_1^1$, $x_2^1$ and $x_3^1$ relative to particle $1$, where particle $1$ and $2$ have configuration space $\R$ and particle $3$ has configuration space $\Z$. Let us write the state $s^1 = (0, n_2^1L+p_2^1, n_3^1L+p_3^1)$, then the state $s^3$ is obtained by $\Lambda^{1 \to 3} \left( T(s^1) \right) = \Lambda^{1 \to 3} (0, n_2^1L,n_3^1L) = (-n_3^1 L, (n_2^1-n_3^1)L,0)$. If for instance $T(x_1^1) = T(x_2^1)$, i.e. $n_1^1=n_2^1$, particle $3$ assigns the same state to particles $1$ and $2$.

\subsubsection{$\R^3 \cong \R^2 \times \R$}
 
Another example is the truncation map $\R^3 \to \R^2$. If we have a symmetry group $\R^2$ in a three-dimensional space, all reference frames along a one-dimensional line are identified with each other. In this case, when applying the truncation map, all group elements in $\R^3$ are projected to the associated group elements in $\R^2$. Consider three systems of which the first two have configuration space $\R^3$ and the last has configuration space $\R^2$. A general state of the systems is $s=(g_1,g_2,g_3)$ where $g_1, g_2 \in \R^3$ and $g_3 \in \R^2$. Relative to particle $1$, the state is $s^1=(e, g_2^1, g_3^1) = (e, n_2^1p_2^1, n_3^1p_C)$ where $p_2^1, p_C \in \R$ and $n_2^1, n_3^1 \in \R^2$. First, we truncate the state: $\tilde{s}=(e=n_1^1, n_2^1, n_3^1)$. Then, we change to the state relative to system $3$: $s^3=(n_1^3, n_2^3,e)$. This can be understood as projecting the points in $\R^3$ to a plane in $\R^2$. If all reference frames along the $z$-axis are identified with each other, this corresponds to projecting on the $x$-$y-$plane.

\subsection{Inconsistency using the truncation method for $G = N \rtimes P$}

In Appendix~\ref{app:imperfect_RF} we prove that the state $s^i$ for a system $i$ with configuration space $N$ obtained from state $s^0$ by truncating and changing reference system is not equivalent to the state $\tilde s^i$ obtained from the state $s$ by first truncating to obtain $\tilde s$, then finding $\tilde s^0$ and then changing reference system to obtain $\tilde s^i$. Let us write the change of reference system $0 \to i$ for a group $G$ as $\Lambda_G^{0 \to i} s^0 = s^i$: 
\begin{align}
\Lambda_G^{0 \to i} (g^0_0, ... , g_{l-1}^0) = \Gamma_R(g^0_i)  (g^0_0, ... , g_{l-1}^0)  \ ,
\end{align}
where $\Gamma_R(g^0_i) s^0$ is $\phi_R(g^0_i, s^0)$, i.e. the right regular action of $g^0_i$ on $s^0$.

\begin{theorem}\label{thm:imperfect_inconsistency}
Let $s = \left((g_0, g_1 , ... , g_{k-1}), (g_{k}, ..., g_{l-1}) \right)$ for $G = N \rtimes P$ with $g_j = n_j$ for $j \in \{k,\ldots,l-1\}$ and $g_i = n_i p_i$ otherwise. Then $\Lambda_N^{0 \to i} \left(T(\REL_G^0(s))\right) \neq   \Lambda_N^{0 \to i} \left(\REL_N^0(T(s))\right)$. Let $\Lambda_N^{0 \to i} \left(T(\REL_G^0(s))\right)  = \left((n_0, n_1 , ... , n_{k-1}), (n_{k}, ..., n_{l-1}) \right)$ and $\Lambda_N^{0 \to i} \left(\REL_N^0(T(s))\right) = \left((m_0, m_1 , ... , m_{k-1}), (m_{k}, ..., m_{l-1}) \right)$, then $n_j = m_j$ for $j \in \{k,\ldots,l-1\}$. For $j \in \{0,\ldots,k-1\}$ this is not always the case.
\end{theorem}  

The above theorem shows that depending on our prior commitment to a well defined state $s$ and our interpretation of relative states $s^i$, the truncation method of obtaining relative states may not be desirable. One may prefer an averaging procedure or a method based on finding invariants. These two approaches are described in Appendix \ref{app:imperfect_RF} using the example of $\SO(3) \rtimes T(\R^3)$. We make use of the truncating procedure in the present work since it can easily be extended to a quantum version using the principle of coherent change of reference system. We describe this quantum generalisation in the next section.

\section{Irreversible changes of quantum reference frame} \label{sec:Irrev_quantum}

The irreversible change of reference frame of the previous section can be extended to the quantum case by applying the principle of \emph{coherent change of reference frame}, as in the reversible case. We describe this in more detail in the following. We note that we only define this change of reference frame for the specific cases $G = N \rtimes P$ and $G= N \times P$.

\subsection{$G = N \rtimes P$ and $G= N \times P$ }

Let us consider the change of reference frame from system $i$ with Hilbert space $L^2(G)$ to a system $j$ with Hilbert space $L^2(N)$. Let us consider the first $k$ systems as being $L^2(G)$ and the last $l-k$ systems as being $L^2(N)$.

A generic product basis state $\ket{\psi^0} = \ket{e}_0  \ket{n^0_1 p^0_1}_1  ... \ket{n^0_{k-1} p^0_{k-1}}_{k-1}   \ket{n^0_{k} p^0_{k}}_k  ...   \ket{n^0_{l-1} p^0_{l-1}}_{l-1}$ (where $p_j^0=e$ for $j \in \{k,\dots, l-1\}$) maps to $\ket{\psi^i} = \ket{e}_i  \ket{n^i_0}_0  \ket{n^i_1}_1  ...   \ket{n^i_{l-1}}_{l-1}$ for $i \in \{k,\dots, l-1\}$. Applying the \emph{coherent change of reference system} we have that the action on a superposition state:
\begin{align}
\ket{\psi^0}  = \ket{e}_0 \bigotimes_{r\neq 0} \ket{n^0_r p^0_r}_r & +  \ket{e}_0 \bigotimes_{s\neq 0} \ket{m^0_s q^0_s}_s \ ,
\end{align}
with $m^0_s \in N$, $q^0_s \in P$ and $p^0_r = p^0_s$ for all $r,s \in \{k,...,l-1\}$ (and similarly for $q^0_s$) maps to:
\begin{align}
\ket{\psi^i} & = \ket{e}_i \bigotimes_{r \neq i} \ket{n^0_r n^i_0}_r +  \ket{e}_i \bigotimes_{s \neq i} \ket{m^0_s m^i_0}_s  \ .
\end{align}

First define the truncation map $\TRUNC: L^2(G) \to L^2(N)$.
\begin{align}
\TRUNC = \int_{n \in N} \int_{p \in P} \ketbra{n}{np} dp dn \ ,
\end{align}
and the map:
\begin{align}
U^{i \to j}_N = &\SWAP_{i,j} \circ  \nonumber \\ &\int_{n^i_j \in N} \ketbra{n^j_i}{n_j^i}_j \otimes \I_i \otimes U_R(n^i_j)^{\otimes l - 2} dn_j^i.
\end{align}
The change of reference frame is given by:
\begin{align}
V^{i \to j} =U^{i \to j}_N \circ \left( \TRUNC^{\otimes k} \otimes \I^{\otimes l-k} \right).
\end{align}

\subsubsection{$\R \cong \Z \rtimes \rU(1)$}\label{sec: example modular}
Consider again the example of the modular truncation of the translation group: $\R \cong \Z \rtimes \rU(1)$. Take three (quantum) systems $\A$, $\B$ and $\C$ where $\A$ has configuration space $\R$ but states relative to $\C$ are encoded in $L\Z$. Changing reference frame from $\A$ to $\C$ requires
\begin{align}
    \ket{0}_\A \ket{x_0}_\B \ket{x_1}_\C & = \ket{0}_\A \ket{n_0L + p_0}_\B \ket{n_1L }_C \nonumber\\
    & \mapsto \ket{0}_\C \ket{-n_1L}_\A \ket{(n_0-n_1)L}_\B.
\end{align}
The truncation operator is:
\begin{align}
\TRUNC = \sum_{n \in \Z} \int_{p \in \rU(1)} dp \ketbra{nL}{nL +p},
\end{align}
and the change of reference frame operator for $L^2(N)$ is:
\begin{align}
U^{\A \to \C}_N = \SWAP_{\A,\C} \circ \int_{n \in N} \ketbra{n^{-1}}{n}_\C \otimes \I_\A \otimes U_R(n)_\B dn.
\end{align}

The following operator performs the change $\A \to \C$:
\begin{align}
    V^{\A \to \C} =   U^{\A \to \C}_N \circ (\I_\A \otimes \TRUNC_\B \otimes \TRUNC_\C).
\end{align}

When applying the change of reference frame from $\A$ to $\C$ to a state in which system $\C$ is in a superposition state relative to $\A$, the state becomes entangled relative to $\C$:
\begin{align}
    &V^{\A \to \C}  \ket{0}_\A \ket{x_0}_\B \left(\ket{x_1}_\C + \ket{x_1'}_\C\right) \nonumber\\
    =& V^{\A \to \C} \ket{0}_\A \ket{n_0L + p_0}_\B \left(\ket{n_1L}_\C + \ket{n_1'L}_\C \right)\nonumber \\
    =& \ket{0}_\C \left(\ket{-n_1L}_\A \ket{(n_0-n_1)L}_\B + \ket{-n_1'L}_\A \ket{(n_0-n_1')L}_\B\right).
\end{align}

However, if $x_1$ and $x_1'$ are located in the same interval relative to $\C$ (with configuration space $L\Z$), the entanglement vanishes. Here, we recognize the dependence of entanglement on the reference frame relative to which it is described. This was already pointed out in \cite{Giacomini_2019}. Note that we observe entanglement only if the uncertainty in the position of $\C$ with respect to $\A$ (i.e. the difference between $x_1$ and $x'_1$) is larger than the resolution $L$ of the configuration space of reference system $\C$.

\subsubsection{$\R^3 \cong \R^2 \times \R$}
 
As mentioned before, the truncation map $\R^3 \to \R^2$ essentially consists of a projection from a three-dimensional configuration space to a two-dimensional one. Here, we project all points in $\R^3$ onto the $x$-$y$-plane.
\begin{align}
    \R^3 &\to \R^2 \nonumber\\
    \vec{P}\coloneqq (x,y,z) &\mapsto \vec{p}\coloneqq (x,y) \ .
\end{align}
On the level of Hilbert spaces, we assign the state $\ket{\vec{P}}$ relative to a three-dimensional configuration space while $\ket{\vec{p}}$ denotes a state in $\mathcal{H}\cong \R^2$. Hence, to change from the state relative to $\A$ with configuration space $\R^3$ to the state relative to $\C$ with configuration space $\R^2$, we apply the operator
\begin{align}
    V^{\A \to \C} = \textrm{SWAP}_{\A,\C} \circ \iint d\vec{P} d\vec{Q} \ketbra{- \vec{p}}{\vec{P}}_\C \otimes \ketbra{\vec{q}- \vec{p}}{\vec{Q}}_\B,
\end{align}
where $\vec{Q}\coloneqq(x',y',z') \mapsto \vec{q}\coloneqq(x',y')$.

\section{Wigner's friend experiment} \label{sec: Wigner's friend experiment}

In this section we consider the Wigner's friend experiment, introduced by E. Wigner \cite{Wigner_1962} in 1961. The thought experiment consists of two observers, Wigner and his friend, and a two-level quantum system. While Wigner is an external observer of the experiment, his friend is located inside an isolated box, together with the system $\S$ in the quantum state $\ket \psi$. Once the experiment is initiated, the friend measures the system in a certain basis. According to the projection postulate of standard quantum theory, the state of the system collapses to one of the eigenstates of the measurement operator. On the other hand, Wigner describes this process from the outside and would assign a unitary evolution. After the measurement of the system by the friend, the outcome of which Wigner does not know, Wigner would assign an entangled state to the joint system of the friend and the system. These two seemingly contradicting prescriptions of standard quantum mechanics are at the core of the so-called Wigner's friend paradox. 

Before proceeding we observe that there is no logical contradiction in the above two descriptions. The state after applying the projection postulate is a state of the system $\S$ alone, of the form $\ket \psi_\S$. The entangled state assigned by Wigner is a state on $\S + \F$ (where $\F$ is the friend) of the form $\ket \phi_{\S\F}$. Since these are states of two different objects ($\S$ versus $\S + \F$) there is no logical contradiction within the postulates of quantum theory. 

We will now apply the relational formalism introduced previously to this apparent paradox. Let us consider $\W$, short for Wigner, describing his friend $\F$ who measures the system $\S$ in the $\{\ket \uparrow, \ket \downarrow\}$ basis. The friend is located inside a perfectly isolated box. We model everything using two-dimensional systems, since we are interested in two degrees of freedom alone. The change of reference frame will therefore be for $\Z_2$, first described in Example~\ref{ex:z2quant}. We consider the ready state of the friend to just be $\ket \uparrow$ and the state of the system before the measurement to be $\ket \psi =\alpha \ket \uparrow + \beta \ket \downarrow$. We first describe the measurement interaction from the point of view of $\W$ starting in the case where $\S$ is in an eigenstate of the measurement operator:
\begin{align}
\ket \uparrow_\F^\W \ket \uparrow_\S^\W &\mapsto \ket \uparrow_\F^\W \ket \uparrow_\S^\W  \ , \\
\ket \uparrow_\F^\W \ket \downarrow_\S^\W &\mapsto \ket \downarrow_\F^\W \ket \downarrow_\S^\W  \ .
\end{align}
The state of the friend depends on the measurement outcome, hence the record of the outcome can be seen as being stored in the state of the friend. The state $\ket{\uparrow}_\F$ is the state `the friend sees up', and similarly for $\ket \downarrow_\F$ and `the friend sees down'.
 
The change of reference frame $\W \mapsto \F$ for the final states gives:
\begin{align}
U^{\W \to \F} \ket \uparrow_\F^\W \ket \uparrow_\S^\W = \ket \uparrow_\W^\F \ket \uparrow_\S^\F \ , \\
U^{\W \to \F} \ket \downarrow_\F^\W \ket \downarrow_\S^\W = \ket \downarrow_\W^\F \ket \uparrow_\S^\F \ , 
\end{align}
where we observe that in both cases the state of $\S$ relative to $\F$ is $\ket \uparrow_\S^\F$. This state encodes the fact that the friend and the system are perfectly correlated in both cases; $\ket \uparrow_\S^\F$ tells us that the state of the friend and the system are related by the identity element. For an arbitrary superposition state of the system the unitary measurement interaction gives the following evolution:
\begin{equation}
\ket{\uparrow}_\F^{\W} (\alpha \ket{\uparrow}_\S^\W + \beta \ket{\downarrow}_\S^\W) \mapsto  \alpha \ket{\uparrow}_\F^{\W}\ket{\uparrow}_\S^{\W} + \beta \ket{\downarrow}_\F^{\W}\ket{\downarrow}_\S^{\W} \ .
\end{equation}

Now, we want to apply the change of reference frame to switch to the perspective of $\F$. If we apply the operator $U^{\W \to \F} $ to the initial state, using the principle of \emph{coherent change of reference system}, we obtain:
\begin{equation}
U^{\W \to \F} \ket{\uparrow}_\F^{\W} (\alpha \ket{\uparrow}_\S^\W + \beta \ket{\downarrow}_\S^\W) = \ket{\uparrow}_\W^\F  (\alpha \ket{\uparrow}_\S^\F + \beta \ket{\downarrow}_\S^\F) \ .
\end{equation}
If we apply the change of reference frame $U^{\W \to \F}$ to the final state we get:
\begin{align} \label{eq: state seen by friend - RF change}
& U^{\W \to \F} \alpha \ket{\uparrow}_\F^{\W}\ket{\uparrow}_\S^{\W} + \beta \ket{\downarrow}_\F^{\W}\ket{\downarrow}_\S^{\W} 
= (\alpha \ket{\uparrow}_\W^{\F} + \beta \ket{\downarrow}_\W^{\F}) \ket{\uparrow}_\S^\F.
\end{align}

If we interpret this result as the state seen by the friend, this would imply that the friend always sees the system as correlated with herself. This is consistent with the initial description from Wigner's perspective, where the state of the friend and system are correlated in both terms of the entangled state $\alpha \ket{\uparrow}_\F^{\W}\ket{\uparrow}_\S^{\W} + \beta \ket{\downarrow}_\F^{\W}\ket{\downarrow}_\S^{\W}$.

Now, we want to compare this result to the state we get if we simply start from the perspective of the friend. She describes the initial state as:
\begin{equation}
\ket{\uparrow}_\W^\F (\alpha \ket{\uparrow}_\S^\F + \beta \ket{\downarrow}_\S^\F) \ ,
\end{equation}
where Wigner is in the ready state. Applying the projection postulate to the system gives:
\begin{align} \label{eq: state seen by friend - projection}
p_\uparrow &= |\alpha|^2: \ket{\uparrow}_\W^\F \ket{\uparrow}_\S^\F \ , \\
p_\downarrow &= |\beta|^2: \ket{\uparrow}_\W^\F \ket{\downarrow}_\S^\F \ .
\end{align}

Whatever outcome the friend observes, she would always describe the system definitely being in either one of the basis states and Wigner in the state $\ket{\uparrow}_\W^\F$. This seems very different to the description we obtain by taking Wigner's perspective and changing to the friend's reference frame. Note however that this does not necessarily imply a contradiction. More precisely, one should be careful when interpreting the state in Equation \eqref{eq: state seen by friend - RF change} as the final state seen by the friend. Rather, one should interpret it as the state Wigner \emph{infers} (concludes) the friend would see. In fact, starting from the state from Wigner's perspective, we only take into account the information that Wigner has at his disposal. Simply changing reference frames by applying a unitary operator can by no means introduce new information, such as the actual outcome observed by the friend.

Hence, it is not a trivial assumption that the state actually seen by the friend should be the same as the state of the system relative to the friend, obtained by changing perspective from Wigner's viewpoint. In fact, this is assumption $\mathcal{C}$ (consistency assumption) of the Frauchiger-Renner no-go theorem \cite{Frauchiger_2018}. In the framework of relational quantum mechanics, one should not assume the consistency condition to hold \emph{a priori}. Indeed in this work we show that what Wigner infers about the friend's state assignments is given by the change of reference frame $U^{\W \to \F}$ and not the seemingly straightforward assumption $\mathcal C$.

We observe that the conclusion that what Wigner can infer about the state of the system relative to the friend is that they are correlated is the same as Rovelli's treatment of the measurement process in relational quantum mechanics~\cite{Rovelli1996}. Here instead of reasoning using Wigner's measurement operators we used an explicit change of reference frame from Wigner to the friend.

One may think that introducing an additional reference system $\sR$ into the box containing $\F$ and $\S$  could help resolve the issue with Wigner's friend. We show in Appendix \ref{app: Wigner with RF} that this is not the case.

\section{Discussion}\label{sec:discussion}

\subsection{Related work} \label{sec: Relation to other approaches}

The underlying approach of this work is based on the relational quantum mechanics of~\cite{Rovelli1996}. Other work formulating a fully relational quantum theory include~\cite{vanrietvelde2018change, hoehn_how_2018, hoehn_switching_2019, Giacomini_2019, castroruiz2019time, vantrietvelde_switching_2018, hoehn2019trinity}, the toy model of~\cite{poulin_toy_2006} and the systematic treatment of quantum reference frames in~\cite{loveridge_relativity_2017,loveridge_symmetry_2018}. The main difference between~\cite{loveridge_relativity_2017,loveridge_symmetry_2018} and the present treatment is that we begin from an explicitly relational state and emphasise the notion of changing between reference frames as opposed to deriving relational states from an external non-relational state. Related approaches to relational quantum mechanics include the perspectival quantum mechanics of~\cite{Bene_perspectival_2001} and the Ithaca interpretation~\cite{mermin_what_1998}.

The notion of quantum reference frames first appeared as part of the debate on the existence of charge superselection rules~\cite{aharonov_quantum_1967}. Later, Aharonov and Kaufherr gave the first explicit study of quantum reference frames \cite{aharonov_quantum_1984}. Typically a description of a quantum system is given relative to a classical measuring device of infinite mass. It was shown that there is a consistent description of quantum systems relative to quantum reference frames of finite mass. This addresses the issue of \emph{universality} of quantum theory; namely quantum systems are usually described relative to an implicit classical reference frame. If quantum theory is universal, then one would expect that reference frames should correspond to quantum systems.  

Given a quantum reference frame $\sR$ and a quantum system $\S$ the relational observables are observables of $\sR + \S$ which are invariant under the symmetry group. Depending on the state of $\sR$ one may recover a relational description of $\S$ which is equivalent to the standard absolute description of $\S$ (i.e. relative to an implicit classical reference frame). Equivalently, for any absolute description of the state or the observables on $\S$, one can find a system $\sR$ such that there is a gauge invariant description on $\S + \sR$. This observation was important for resolving the `optical coherence controversy' \cite{bartlett_dialogue_2006}. Explicit maps between the absolute and relativised descriptions can be found in~\cite{Bartlett_2007,loveridge_symmetry_2018}, where an emphasis on the localisation of the reference system is placed in~\cite{loveridge_relativity_2017,loveridge_symmetry_2018}. Other works discussing relational observables include \cite{vanrietvelde2018change, hoehn_how_2018, hoehn_switching_2019, vantrietvelde_switching_2018, hoehn2019trinity}.

In~\cite{Angelo_2011,Angelo_2012} a description of quantum systems relative to other quantum systems is given, in particular for the translation group. The initial description of $\sR$ and $\S$ is given relative to an external classical reference frame (in the position basis), and the description relative to $\sR$ is obtained by refactoring into center of mass and relative position. Tracing out the center of mass partition then allows to remove all global degrees of freedom and leaves us with a relational description of the systems. Moreover the description of systems is shown to be dependent on the mass of the quantum reference system. Note that such global degrees of freedom never enter into our formalism in the first place. Instead, we give the states of systems relative to a specified system from the outset.

The change of reference frame in this work generalises the known changes of reference frame $L^2(\R)$ and $L^2(\R^3)$ for spatial position of~\cite{Giacomini_2019,vantrietvelde_switching_2018}, and $L^2(\R)$ for temporal degrees of freedom of~\cite{hoehn_how_2018,hoehn_switching_2019,castroruiz2019time,hoehn2019trinity}. Reference frames for rotational degrees of freedom are studied in the perspective-neutral approach in~\cite{vantrietvelde_switching_2018}. Brukner and Mikusch are independently working on a treatment of rotational degrees of freedom of quantum reference frames using large spin coherent states.

We observe that in both~\cite{Giacomini_2019} and the present work the consistency of the fully relational account with a description which is initially given externally (and from which the relative description is then obtained) is not proven. However in~\cite{vanrietvelde2018change} it is shown that there is a `perspective-neutral' framework which encompasses all perspectives for~\cite{Giacomini_2019}. Note that this framework does not describe an external structure as it only encodes relative information. Extending this perspective-neutral approach to the general cases in the present paper could constitute an interesting direction for future work.

In~\cite{zych_relativity_2018} an emphasis is placed on the notion that a reference system in a superposition relative to another system gives new coordinates, which are not related by a classical coordinate transformation to the initial ones. This is conceptually closer to the perspective in the present work than that of~\cite{Giacomini_2019}. Namely we do not make the claim that the relative descriptions are `operational' as in~\cite{Giacomini_2019}.

Prior work with an emphasis on changing perspectives between quantum reference frames is found in~\cite{Palmer_2014} where this change of reference frame is mediated by an external description, and uses tools such as $G$-twirling.

In recent years there has been a significant interest in the study of quantum reference frames as resources for measurements, communication tasks and thermodynamic exchanges amongst others~\cite{gour_resource_2008,popescu_quantum_2018,poulin_dynamics_2007,bartlett_quantum_2009,haplern_microcanonical_2016,miyadera_approximating_2016,skotiniotis_macroscopic_2017}. This approach is not motivated by the same considerations as the present paper and the link between the two approaches is not fully clear to the authors. It is possible that the present exposition, carried out explicitly in a group theoretic language, could be used to relate the two. For instance the `perfect reference frames' of \cite{Bartlett_2007} are of the form $L^2(G)$, which also plays a prominent role here. The link between imperfect reference frames as standardly defined~\cite{Bartlett_2007,bartlett_quantum_2009,miyadera_approximating_2016} and the truncation based approach of the present work also remains to be worked out.

\subsection{General comments}

\subsubsection{Relational quantum mechanics}

We have provided a formalism for relational quantum mechanics which captures its spirit and recovers the same conclusions for Wigner's friend and the Frauchiger-Renner theorem. Future work could involve applying the formalism developed in this work to other thought experiments, in order to provide an explicit account of how relational quantum mechanics addresses various apparent paradoxes. 

\subsubsection{Observer dependence of the symmetry group}

One further contribution of this work is that symmetries are not only relative to the system being described, but also to the system doing the describing. A system with Hilbert space $\H$ has a symmetry group $G$ if it carries a representation of $G$, however it can only be described as transforming under $G$ by a system which carries a regular representation of $G$. For example although a qubit carries a representation of $\SU(2)$, two qubits `describing' each other and a third qubit would use the subgroup $\Z_2$, since they can only carry a regular representation of $\Z_2$ and not the full $\SU(2)$. This is reminiscent of the argument by Penrose in \cite{Penrose}.

\subsubsection{$G = N \rtimes P$ and $G = N \times P$}

For the imperfect reference frames we describe a very specific case, though it seems natural since it is the structure of well known space time groups. Group extension tells us that for any group subgroup pair $(G,N)$ with $N$ normal, the extension is of the form $G = N \rtimes P$ or $G = N \times P$.  Hence our approach fully covers imperfect reference frames for group subgroup pairs $(G,N)$ with $N$ normal. However for a full account of imperfect reference frames one would need to relax the assumption of $N$ normal and see whether one can define meaningful changes of reference frame.

\section{Conclusion} \label{sec: conclusion}

The construction of a relational quantum theory requires a description of systems and states relative to other systems. In this work, we present a formalism that allows to describe the relative states of systems and to change between the descriptions of different reference systems. Starting from the analysis of reference frames in the classical realm, we moved on to the description of quantum states relative to quantum reference frames. Depending on the system whose viewpoint is adopted, the description of physical phenomena changes. We find that quantum properties such as entanglement and superposition are not absolute but depend on the reference frame relative to which they are described. As such the conclusions of~\cite{Giacomini_2019} are shown to be generic to quantum reference frames for arbitrary groups.

Previous work on quantum reference frames in the relational paradigm restricted itself to the translation group (for time, space and momentum translations) and the Euclidean group in three dimensions (including rotations)~\cite{vantrietvelde_switching_2018}. Here we have extended this to arbitrary groups, including finite groups and non-Abelian groups. Moreover we have highlighted the key representation theoretic features needed for reversible changes of quantum reference frame of a group $G$, namely that the quantum reference systems have Hilbert space $L^2(G)$. Using this insight we have developed a more abstract formulation directly in terms of the relevant symmetry group. 

We also extended the study of reference frames in the relational approach to imperfect reference frames using a `truncation' approach, in contrast to group averaging approaches more common in quantum information based approaches.
Future work involves exploring imperfect reference frames for different cases than $(G,N)$ with $N$ a normal subgroup.
One benefit of our approach is that it formulates the relational approach of~\cite{Giacomini_2019} in a group theoretic language which is more common in the quantum information literature. As such it may help in developing a fully rigorous account of the link between the two.  Recent work \cite{hoehn2019trinity} in this direction has been exploring links between different approaches in the case of the translation group and would be useful to pursue in this more general case.

Our exploration of the Wigner's friend scenario shows how to apply our relational framework to an important discussion point in the literature. Using our approach we reach the same conclusion as Rovelli in his relational treatment of the measurement process in~\cite{Rovelli1996}: all that can be said by Wigner is that the friend is correlated with the measurement outcome. The novel aspect of this work is the use of an explicitly relational formalism (which embodies the philosophical position of relational quantum mechanics) as well as obtaining the friend's perspective through an explicit change of reference frame, as opposed to reasoning indirectly using measurement operators.

\begin{acknowledgments}
The authors thank \v{C}aslav Brukner, Esteban Castro-Ruiz, Flaminia Giacomini, Matt Leifer, Leon Loveridge, Pierre Martin-Dussaud, Carlo Rovelli, David Schmid and Rob Spekkens for helpful discussions. The authors thank Philipp H{\" o}hn, Leon Loveridge and Markus M{\"u}ller for helpful comments on a draft of this paper.
This research was supported by Perimeter Institute for Theoretical Physics. Research at Perimeter Institute is supported in part by the Government of Canada through the Department of Innovation, Science and Economic Development Canada and by the Province of Ontario through the Ministry of Colleges and Universities.
\end{acknowledgments}

\nocite{apsrev42Control}
\bibliographystyle{apsrev4-2.bst}
\bibliography{mybibliography} % Produces the bibliography via BibTeX.

\appendix

\section{Background}\label{app:background}

\subsection{Coordinate systems and coordinate charts}

We distinguish two separate notions: coordinate systems and coordinate charts following the presentation in~\cite{baez_dimensional_2006}. A \emph{coordinate system} is an isomorphism $f: X \to Y$ where $X$ is the object of interest and $Y$ is some known object we are using to describe $X$. In the terminology of Korzybski, $X$ is the territory and $Y$ is the map. 

However it is often the case (as in general relativity) that there is no isomorphism between the object (space-time) and the description of the object (coordinate chart). A \emph{coordinate chart} is a map in the opposite direction: $f: Y \to X$ where once more $X$ is the object of interest and $Y$ is the known object being used to describe $X$, where the map $f$ is no longer an isomorphism (for instance for $X$ a curved manifold one needs multiple coordinate charts $ \cong \R^n$ to describe $X$ fully).

In the present paper we consider coordinate systems of a manifold $X$ and not coordinate charts.

\subsection{Coordinate systems on G-torsors}

Given a space $X \cong G$ for some group $G$ it is clear that every $g \in G$ gives rise to an isomorphism $f: X \to Y$ (where $Y \cong G$) via the map $f: x \mapsto gx$. While $e$ is the identity element of the group $G$ it is the \emph{origin} of the $G$-torsor. Different isomorphisms $f: X \to Y$ correspond to different choices of origin. To quote~\cite{baez_torsors_2009}: `A torsor is like a group that has forgotten its identity.' Since every $g \in G$ leads to a different coordinate system, the set of different choices of coordinate systems is also $G$. As noted in the main body in the case where $X \cong G$ is a Lie group and we are not concerned with the group structure but only the smooth differentiable manifold structure one could find other coordinate systems via the isomorphism $f: x \mapsto  mx$ where $m$ is an element of $\Diff(X)$, the diffeomorphism group on $X$. In the case where $X \cong \R$ this would consist of considering coordinate systems which are related by more general transformations than translations, for instance re-scaling by a real scalar.

\subsection{Reference frames}

A \emph{reference frame} is a coordinate system together with a physical system whose configuration uniquely determines the coordinate system. Since every choice of coordinate system is in correspondence with an element $g \in G$ for a space $X \cong G$, a valid choice of physical system used to form a reference frame is a physical system with configuration space $G$. 

\section{Proofs of Lemmas~\ref{lem:unit},~\ref{lem:inverse} and~\ref{lem:transitive}} \label{app:proofs_of_lems}

\subsection{Proof of Lemma~\ref{lem:unit}}

We show that $U^{i \to j}$ is unitary.
%(see Equation~\eqref{eq:change_of_RF}). 

To change reference frame between systems $i$ and $j$, where $i$ and $j$ have state space $L^2(G)$, and $n-2$ other systems carrying the right regular representation $U_k(g_j^i)\ket{\psi(g_k^i)}_k = \ket{\psi(g_k^ig_i^j)}_k=\ket{\psi(g_k^j)}_k$, the general operator is 
\begin{align}
    U^{i \to j} = \SWAP_{i,j} \circ \int_{g^i_j \in G} dg_j^i\  \ketbra{g^j_i}{g_j^i}_j \otimes \I_i \otimes \nonumber \\ \bigotimes_{k\neq i,j} U_k(g_j^i).
\end{align}
We can show that this operator is unitary:
\begin{align}
    (U^{i \to j})^\dagger = \int_{h^i_j \in G} dh_j^i\  \ketbra{h^i_j}{h_i^j}_j \otimes \I_i \otimes \nonumber \\ \bigotimes_{k\neq i,j} U^\dagger_k(h_j^i) \circ \SWAP_{i,j}^\dagger.
\end{align}
Hence,
\begingroup
\allowdisplaybreaks
\begin{align*}
    &(U^{i \to j})^\dagger U^{i \to j} \\
= &   \int_{h^i_j \in G} dh_j^i\   \ketbra{h^i_j}{h_i^j}_j \otimes \I_i \otimes \bigotimes_{k\neq i,j} U^\dagger_k(h_j^i) \circ \SWAP_{i,j}^\dagger \\
& \circ \SWAP_{i,j} \circ \int_{g^i_j \in G} dg_j^i\  \ketbra{g^j_i}{g_j^i}_j \otimes \I_i \otimes \bigotimes_{k\neq i,j} U_k(g_j^i) \\
= & \int_{h^i_j \in G} dh_j^i \int_{g^i_j \in G} dg_j^i\  \ket{h^i_j}\braket{h_i^j}{g^j_i}\bra{g_j^i}_j \otimes \I_i\  \otimes \\
&\bigotimes_{k\neq i,j} U^\dagger_k(h_j^i)U_k(g_j^i) \\
& = \int_{g^i_j \in G} dg_j^i\ \ketbra{g^i_j}{g_j^i}_j \otimes \I_i \otimes \bigotimes_{k\neq i,j} U^\dagger_k(g_j^i)U_k(g_j^i) \\
& =\I_j \otimes \I_i \otimes \bigotimes_{k\neq i,j} U_k((g_j^i)^{-1}g_j^i) \\
& =\I_j \otimes \I_i \otimes \bigotimes_{k\neq i,j} \I_k \\
& = \I.
\end{align*}
\endgroup
Here, we used $\braket{h_i^j}{g_i^j}=\delta_{gh}$, $\int dg_j^i\ \ketbra{g^i_j}{g_j^i}_j = \I_j$ and $U_k(e)=\I_k$.

\subsection{Proof of Lemma~\ref{lem:inverse}}

We show $\left(U^{0 \to i}\right)^\dagger = U^{i \to 0}$.

\begin{align*}
&\left(U^{0 \to i}\right)^\dagger \\
& = \int_{g^0_i \in G}\ dg_i^0 \ketbra{g_i^0}{g^i_0}_i \otimes \I_0 \otimes U_R^\dagger(g^0_i)^{\otimes n - 2} \circ \SWAP_{0,i}^\dagger \\
& = \SWAP_{i,0} \circ \int_{g^0_i \in G}\ dg_i^0 \ketbra{g_i^0}{g^i_0}_0 \otimes \I_i \otimes U_R^\dagger(g^0_i)^{\otimes n - 2} \\
& = \SWAP_{i,0} \circ \int_{g^i_0 \in G}\ dg_0^i \ketbra{g_i^0}{g^i_0}_0 \otimes \I_i \otimes U_R(g^i_0)^{\otimes n - 2} \\
& = U^{i \to 0}
\end{align*}
From the second to the third line, we commuted the SWAP operator through to the left by changing the labels of partitions $0$ and $i$. From the third to the fourth line, we used the fact that the integral over $g_i^0$ is the same as the integral over $g_0^i$. Moreover, it holds that $U_R^\dagger(g^0_i) = U_R(g^i_0)$ because $U_R(g_0^i)U_R(g_i^0)\ket{g}=\ket{gg_0^i g_i^0} = \ket{g}$, hence $U_R(g_0^i)=U_R^\dagger(g_i^0)$.

\subsection{Proof of Lemma~\ref{lem:transitive}}
We prove that $U^{i \to j} U^{k \to i} = U^{k \to j}$.

For this, we show how the operators act on an arbitrary basis state:
\begin{align*}
    &U^{i \to j} U^{k \to i} \ket{e}_k \ket{g^k_0}_0 ... \ket{g^k_i}_i ...\ket{g^k_j}_j ... \ket{ g_{n-1}^k}_{n-1}\\
    &=\!U^{i \to j}\! \SWAP_{k,i}\! \ket{e}_k\! \ket{g^k_0\!g_k^i}_0\! ...\! \ket{g^k_i}_i \!...\!\ket{g^k_j\!g_k^i}_j\! ...\!\ket{ g_{n-1}^k\!g_k^i}_{n-1} \\
    &=U^{i \to j} \ket{e}_i \ket{g^k_0g_k^i}_0 ... \ket{g^k_i}_k ...\ket{g^k_jg_k^i}_j ... \ket{ g_{n-1}^kg_k^i}_{n-1} \\ 
    &=\SWAP_{i,j} \ket{e}_i \ket{g_0^ig_i^j}_0\! ...\! \ket{g^k_ig_i^j}_k\! ...\!\ket{g_i^j}_j\! ...\! \ket{ g_{n-1}^ig_i^j}_{n-1} \\
    &= \ket{e}_j \ket{g_0^j}_0 ... \ket{g_k^j}_k ...\ket{g_i^j}_i ... \ket{ g_{n-1}^j}_{n-1} \\
    &=U^{k \to j} \ket{e}_k \ket{g^k_0}_0 ... \ket{g^k_i}_i ...\ket{g^k_j}_j ... \ket{ g_{n-1}^k}_{n-1}.
\end{align*}

\section{On the unitarity of the change of reference frame operator} \label{app: unitarity of RF change operator}

Let us consider  $\ket{\psi(g)}$, with $U_L(k)\ket{\psi(g)} = \ket{\psi(kg)}$. Take two initial states defined relative to $\A$:
\begin{align}
\ket{\psi^1} = \ket{\psi(e)}_\A \ket{\psi(g_1)}_\B \ket{\psi(g_2)}_\C \ ,\\
\ket{\psi^2} = \ket{\psi(e)}_\A \ket{\psi(h_1)}_\B \ket{\psi(g_2)}_\C.
\end{align}

The change of reference frame $\A \to \B$: $\ket{\psi^i} \mapsto \ket{\phi^i}$ gives:
\begin{align}
\ket{\phi^1} = \ket{\psi(e)}_\B \ket{\psi(g_1^{-1})}_\A \ket{\psi(g_2g_1^{-1})}_\C \ ,\\
\ket{\phi^2} = \ket{\psi(e)}_\B \ket{\psi(h_1^{-1})}_\A \ket{\psi(g_2h_1^{-1})}_\C.
\end{align}

For a superposition state $\ket{\psi^3} = \alpha \ket{\psi^1} + \beta \ket{\psi^2}$ ($\alpha, \beta \neq 0$) the change of reference system gives:
\begin{align}
& \ket{\psi^3} = \alpha \ket{\psi^1} + \beta \ket{\psi^2} \mapsto \ket{\phi^3} = \alpha \ket{\phi^1} + \beta \ket{\phi^2}  \ ,
\end{align}
where 
\begin{align}
\ket{\phi^3} & =  \ket{\psi(e)}_\B  ( \alpha \ket{\psi(g_1^{-1})}_\A \ket{\psi(g_2g_1^{-1})}_\C  \nonumber\\
&+ \beta  \ket{\psi(h_1^{-1})}_\A \ket{\psi(g_2h_1^{-1})}_\C  ) \ .
\end{align}

Let us show that unitarity of the change of reference frame implies that the states $\ket{\psi(g)}$ are mutually orthogonal. We compute the inner product of two initial states $\ket{\psi^1}$ and $\ket{\psi^3}$:
\begin{align}
\braket{\psi^1}{\psi^3} = \alpha + \beta \braket{\psi(g_1)}{\psi(h_1)}_\B \ ,
\end{align}
and the inner product of the two final states $\ket{\phi^1}$ and $\ket{\phi^3}$:
\begin{align}
\braket{\phi^1}{\phi^3} = \alpha + \beta \braket{\phi^1}{\phi^2} \ , 
\end{align}
where 
\begin{align}
\braket{\phi^1}{\phi^2} = \braket{\psi(g_1^{-1})}{\psi(h_1^{-1})}_\A \braket{\psi(g_2g_1^{-1})}{\psi(g_2h_1^{-1})}_\C  \ .
\end{align}
Since $\ket{\psi(g_2g_1^{-1})} = U(g_2) \ket{\psi(g_1^{-1})}$ and $\ket{\psi(g_2h_1^{-1})} = U(g_2) \ket{\psi(h_1^{-1})}$ we have that $\braket{\psi(g_2g_1^{-1})}{\psi(g_2h_1^{-1})}_\C = \braket{\psi(g_1^{-1})}{\psi(h_1^{-1})}_\C$.  
Therefore:
\begin{align}
\braket{\phi^1}{\phi^3} = \alpha + \beta ( \braket{\psi(g_1^{-1})}{\psi(h_1^{-1})}^2).
\end{align}

By assumption there are two commuting actions: $U_L(g) \ket{\psi(h)} = \ket{\psi(gh)}$ and $U_R(g)\ket{\psi(h)} =  \ket{\psi(h g^{-1})}$. 

%Now $\braket{\psi^1}{\psi^3} = \braket{\phi^1}{\phi^3}$ requires $ \braket{\psi(g_1^{-1})}{\psi(h_1^{-1})} =  \braket{\psi(g_1)}{\psi(h_1)}^2$. We assume this holds and show it implies that $\braket{\psi(g)}{\psi(h)} = \delta(h,g)$.
Now $\braket{\psi^1}{\psi^3} = \braket{\phi^1}{\phi^3}$ requires $ \braket{\psi(g_1)}{\psi(h_1)} =  \braket{\psi(g_1^{-1})}{\psi(h_1^{-1})}^2$. We assume this holds and show it implies that $\braket{\psi(g)}{\psi(h)} = \delta(h,g)$.

\begin{align*}
& \braket{\psi(g_1)}{\psi(h_1)} =  \braket{\psi(g_1^{-1})}{\psi(h_1^{-1})}^2 \\
& \Leftrightarrow \braket{\psi(g_1)}{\psi(h_1)} =  \overline{\braket{\psi(g_1^{-1})}{\psi(h_1^{-1})}}^2 \nonumber\\
& \Leftrightarrow  \braket {U_L(g_1^{-1}) \psi(g_1)}{U_L(g_1^{-1}) \psi(h_1)} \nonumber\\
&=   \overline{\braket{U_R(h_1^{-1}) \psi(h_1^{-1})}{U_R(h_1^{-1}) \psi(g_1^{-1})}}^2 \nonumber\\
& \Leftrightarrow  \braket{ \psi(e)}{ \psi(g_1^{-1} h_1)} =  \overline{\braket{\psi(e)}{\psi(g_1^{-1} h_1)}}^2 
\end{align*}

Which only holds when $\ket{\psi(g_1^{-1} h_1)}$ is orthogonal to $\ket{\psi(e)}$ (when $g_1 \neq h_1$). Therefore for all $g \in G$ it is the case that $\braket{\psi(e)}{\psi(g)} = \delta(g,e)$.

Now consider $\braket{\psi(g)}{\psi(h)}$ for arbitrary $g,h \in G$. This is equal to $\braket{U_L(g^{-1})\psi(g)}{U_L(g^{-1})\psi(h)} = \braket{\psi(e)}{\psi(g^{-1}h)} = \delta(e,g^{-1}h) = \delta(g,h)$. This entails that the states $\ket{\psi(g)}$ form an orthonormal basis for $\H_i$ or a subspace of it if they do not span the full space.

Hence, the operator which performs the required change of reference frame and obeys the principle of coherent change of reference frame is unitary if and only if the states $\ket{\psi(g)}$ form an orthonormal basis for $\H_i$ or a subspace thereof.

\section{Imperfect reference frames}\label{app:imperfect_RF}

\subsection{Proof of Theorem~\ref{thm:imperfect_inconsistency}}
\begin{align}
s &= \left((g_0,...,g_{k-1}),(g_{k},...,g_{l-1}) \right) \\
&= \left((n_0 p_0,...,n_{k-1} p_{k-1}),(n_{k},...,n_{l-1})\right)
\end{align}
We first find the state $\tilde s^j = \Lambda^{0 \to j}\left(\REL_N^0(T(s))\right)$.
Let us truncate $s$:
\begin{align}
\tilde s =  T(s) = \left((n_0 ,...,n_{k-1} ),(n_{k},...,n_{l-1})\right).
\end{align}
Then the state $\tilde s^0$ is:
\begin{align}
\tilde s^0= \REL_N^0  \left((n_0^0 ,...,n_{k-1}^0 ),(n_{k}^0,...,n_{l-1}^0)\right).
\end{align}
Then the state $\tilde s^j = \Lambda^{0 \to j} \tilde s^0$ is:
\begin{align}
T(s)^j & = \left((n_0^0 n_0^j ,...,n_{k-1}^0 n_0^j ),(n_{k}^0 n_0^j,...,n_{l-1}^0 n_0^j)\right) \\
& = \left((n_0^j ,...,n_{k-1}^j ),(n_{k}^j,...,n_{l-1}^j)\right)
\end{align}
where $n^i_j n_i = n_j$ and $n^r_i n_r^j = n_i^j$.

Now we compute $\tilde s^j = \Lambda^{0 \to j}\left(T(\REL_G^0 (s))\right)$. $s^0 = \REL_G^0 s$ is:
\begin{align}
s^0 &= \left((g_0^0,...,g_{k-1}^0),(g_{k}^0,...,g_{l-1}^0) \right) \\
& = \left((m_0^0 q_0^0,...,m_{k-1}^0 q_{k-1}^0),(m_{k}^0 q_{k}^0,...,m_{l-1}^0 q_{l-1}^0)\right) \ ,
\end{align}
where $g_i^j = m_i^j q_i^j$ is the decomposition into $NP$. Let us consider the elements $m_j^0 q_j^0$ for $j \in \{k,\dots, l-1\}$. We have that $g^0_j g_0 = n_j \rightarrow m_j^0 q_j^0  n_0 p_0 = n_j$. Now observe that since $m_j^0$ and $n_j$ in $N$ this implies that $q_j^0  n_0 p_0 = n' \in N$. Now there is a unique  $q_j^0$ such that this holds, and observe that $p_0^{-1} n_0 p_0 = m' $. There is a unique $m \in N$ such that $mm'=n'$. Therefore $q_j^0 =mp_0^{-1}$. However since $q_j^0 \in P$ this implies $m = e$ and therefore  $q_j^0 = p_0^{-1}$.

Let us truncate:
\begin{align}
T(s^0) = \left((m_0^0 ,...,m_{k-1}^0 ),(m_{k}^0,...,m_{l-1}^0)\right).
\end{align} 
Then one can obtain $T(s^0)^j = \Lambda^{0 \to j}T(s^0) = \Gamma_R(m_{j}^0) T(s^0)$:
\begin{align}
T(s^0)^j = \left((m_{0}^j ,...,m_{k-1}^0 m_{0}^j ),(m_{k}^0 m_{0}^j,...,m_{l-1}^0 m_{0}^j)\right).
\end{align} 
Does $m_i^0 m_{0}^j = n_{i}^j$, i.e. does $m_i^0 m_{0}^j n_j = n_i$. There are two cases: $i \in \{0,\ldots, k-1\}$ and $i \in \{k,\ldots, l-1\}$. Let us consider the second:
\begin{align}
& m^0_i q^0_i n_0 p_0 = n_i \\
& m^0_i = n_i (p_0^{-1} n_0 p_0)^{-1}
\end{align}
since $q^0_i = p_0^{-1}$. Similarly $m^0_j = n_j (p_0^{-1} n_0 p_0)^{-1}$. Therefore
\begin{align}
m_i^0 m_{0}^j = m_i^0 (m_{0}^j)^{-1} &= n_i (p_0^{-1} n_0 p_0)^{-1}  (p_0^{-1} n_0 p_0) n_j^{-1} \ , \nonumber \\
&= n_i n_j^{-1}.
\end{align}
This is indeed $n_{i}^j$ since it maps $n_j$ to $n_i$.

Now let us consider the case $i \in \{0,\dots, k-1\}$:
\begin{align}
m^0_i q^0_i n_0 p_0 = n_i p_i
\end{align}
where $q^0_i$ is not necessarily equal to $p_0^{-1}$.
Then we obtain 
\begin{align}
m_i^0 m_{0}^j & = m_i^0  (m_{0}^j)^{-1} = n_i p_i (q^0_i n_0 p_0)^{-1}  (p_0^{-1} n_0 p_0) n_j^{-1} \\
& = n_i p_i  p_0^{-1}n_0^{-1} (q^0_i)^{-1} p_0^{-1} n_0 p_0 n_j^{-1}
\end{align}
which is not equal to $n_{i}^j$ in general. For a specific example we can look at $\R = \Z \rtimes S^1$. Consider
\begin{align}
s = (n_0 L + x_0, n_1 L + x_1, n_2 L , n_3 L ).
\end{align} 
Then $T(s) = (n_0 L, n_1 L, n_2 L, n_3L)$ and $T(s)^2=(n_0 - n_2, n_1 -n_2, 0, n_3-n_2)L$. 
For the other order we get $s^0 = (0 , n_1 L + x_1 - (n_0 L + x_0),  n_2 L  - (n_0 L + x_0),n_3 L  - (n_0 L + x_0))$. Let us assume that $x_1 - x_0 < 0$, then $T(s^0) = (0,n_1 - n_0 -1 , n_2 - n_0, n_3- n_0)L$ and $T(s^0)^2 = (-(n_2 - n_0), n_1 - n_0 -1-(n_2 - n_0) , n_2 - n_0-(n_2 - n_0), n_3- n_0-(n_2 - n_0))L$ which gives $T(s^0)^2 = (n_0 - n_2,n_1 -n_2 - 1, 0, n_3-n_2)L$ and does not equal $T(s)^2$.

\subsection{Averaging and invariants for the example of $\SO(3) \rtimes T(\R^3)$} \label{app: subsection averaging}

Now let us return to the case where the configuration space is $\R^3$ but the transformation group is $E^+(3) \cong T(\R^3) \rtimes \SO(3)$ and the case of three particles. Here $E^+(3)$ is the special Euclidean group consisting of translations followed by a rotation. Let us say that particle $0$ uses a coordinate system $(x^0, y^0, z^0)$ such that it assigns itself the state $(0,0,0) = \vec 0$. There are infinitely many such coordinate systems corresponding to all reference frames centred on particle $0$, related by rotations $O \in \SO(3)$. We write $\vec v^0_i =(x^0_i, y^0_i, z^0_i)$ for the state of particle $i$ in a reference frame centred at $0$. The coordinates corresponding to Cartesian reference frames are in one to one correspondence with group elements in  $T(\R^3) \rtimes \SO(3)$. Each state $\vec v^0_i =(x^0_i, y^0_i, z^0_i)$ is stabilized by a $\SO(3)$ subgroup: $\vec v^0_i \rtimes O$ $\forall O \in \SO(3)$ . We can write $\R^3 \cong E^+(3) /\SO(3)$.

The description of the state of the three particles is $s^0_{0,1,2} = \left( \vec 0, \vec v^0_1, \vec v^0_2\right)$. What can particle $0$ infer about the description used by particle 1, knowing only that the convention is such that particle 1 uses Cartesian coordinates placed at its position? There are infinitely many such choices of coordinates, related to the coordinates $(x^0_i, y^0_i, z^0_i)$ by $t_{-v^0_1} \rtimes O$ for all $O \in \SO(3)$. Hence the states  $s^{1,O}_{0,1,2} = \left (-O\vec{v^0_1}, \vec 0, O\left(\vec v^0_2-  \vec{v^0_1} \right)\right)$ for any $O \in \SO(3)$ correspond to coordinate choices centred on particle $1$. In order to describe the ignorance about $O$ one needs a probabilistic representation of states: a state is now a measure on $\R^3$. The state $-\vec{a}$ is the Dirac measure $\delta_{-\vec{a}}$ and the description that particle 0 assigns to particle 1 is:
\begin{align}
s^{1, av}_{0,1,2} &  = \int_{O \in \SO(3)} \left(O \delta_{-\vec{v^0_1}} , \delta_{\vec{0}}, \delta_{\vec v^0_2-  \vec{v^0_1}}\right) \\
&= \left(\mu_{S^2, |\vec{v^0_1}|}, \delta_{\vec{0}},\mu_{S^2, |\vec v^0_2-  \vec{v^0_1}|}\right) \ ,
\end{align}
where $\mu_{S^2, |\vec{a}|}$ is the normalised Haar measure on the sphere $S^2$ of radius $|\vec a|$ centred at the origin (in particle $1$'s coordinates). We observe that there is no reversible transformation from $s^{1, \av}_{0,1}$ representing particle $0$'s knowledge of particle $1$'s description back to the initial description of particle 0. 
Starting from $s^0$ and mapping to $s^2$ one obtains:
\begin{align}
s^{2, av}_{0,1,2} &  = \int_{O \in \SO(3)} \left(O \delta_{-\vec v^0_2} , \delta_{\vec v^0_1 -  \vec{v}^0_2}, \delta_{0}\right) \\
&= \left(\mu_{S^2, |\vec{v^0_2}|},\mu_{S^2, |\vec v^0_1-  \vec{v^0_2}|}, \delta_{\vec{0}}\right) .
\end{align}
We leave to future work the development of a full account of changes of reference frame in this probabilistic case (for instance it is not clear that there is a well defined change $s^{1, av}_{0,1,2} \mapsto s^{2, av}_{0,1,2}$). Here we are interested only in the single change from $s^0 \mapsto s^1$.

We also note that when describing ignorance one cannot integrate over the pure states, or configurations. In the above example this would result in $s^{1, \av}_{0,1}$ being $\left( \vec 0 , \vec 0\right)$ which is nonsensical. This is similar to the case in quantum theory where integrating over a group action on pure states does not represent ignorance (rather it is a projection) but integrating over mixed states corresponds to ignorance. One must move to a probabilistic description of states in terms of measures on $X$ in order to infer states of other particles for these types of cases. Alternatively one could stay at the pure state description, and instead of finding the point of view that particle 0 infers particle 1 holds, we can search for what particle 0 knows holds true for all possible descriptions  $s^1_{0,1} = \left (O\vec{-a}, \vec 0 \right)$ for all $O \in \SO(3)$ from the point of view of particle 1. In other words: what are the \emph{invariants}? Here it is straightforward to see that any $f(|\vec{a}|)$ is an invariant quantity.

\section{Wigner's friend with additional reference system} \label{app: Wigner with RF}

One could argue that introducing an additional reference system $\sR$ into the box containing $\F$ and $\S$ might help to resolve the paradox of Wigner's friend. 
In this case, we can describe the measurement interaction from the point of view of $\W$ as follows:
\begin{equation}
\ket{\uparrow}_\sR^{\W}\ket{\uparrow}_\F^{\W}\ket{\psi}_\S^{\W} \to  \ket{\uparrow}_\sR^{\W} (\alpha \ket{\uparrow}_\F^{\W}\ket{\uparrow}_\S^{\W} + \beta \ket{\downarrow}_\F^{\W}\ket{\downarrow}_\S^{\W}).
\end{equation}
The reference system is not affected by the measurement.
Now, we want to apply the change of reference frame to switch to the perspective of $\F$. We apply the operator $U^{\W \to \F}_{\A \S}$ to the initial state:
\begin{align}
U^{\W \to \F}_{\sR \S} \ket{\uparrow}_\sR^{\W}\ket{\uparrow}_\F^{\W}\ket{\psi}_\S^{\W} = \ket{\uparrow}_\W^\F \ket{\uparrow}_\sR^\F \ket{\psi}_\S^{\F}\ , 
\end{align}
and the final state:
\begin{align}
&U^{\W \to \F}_{\sR \S} \ket{\uparrow}_\sR^{\W} (\alpha \ket{\uparrow}_\F^{\W}\ket{\uparrow}_\S^{\W} + \beta \ket{\downarrow}_\F^{\W}\ket{\downarrow}_\S^{\W}) \nonumber\\ &= (\alpha \ket{\uparrow}_\W^{\F}\ket{\uparrow}_\sR^{\F} + \beta \ket{\downarrow}_\W^{\F}\ket{\downarrow}_\sR^{\F}) \ket{\uparrow}_\S^\F.
\end{align}
We see that, again, the state of the system is always correlated with the state of the friend. Wigner and the additional reference system end up being entangled. We conclude that introducing an additional reference system into the laboratory does not provide a resolution to the paradox.

\section{Comparison to RF change operator of \cite{Giacomini_2019}} \label{app: our operator vs Giacomini operator}
This is the formal proof that the operator given in \cite{Giacomini_2019} for the change between two reference frames $\A$ and $\B$ on the real line (one-dimensional translation group) is equivalent to the operator given in Equation \eqref{eq: RF change translation group}, namely 
\begin{align} \label{eq: app: RF change translation group}
  U^{\A \to \B} = \SWAP_{\A,\B} \circ \int dx dy\ \ketbra{-x}{x}_\B \otimes \mathbb{1}_A \otimes \ketbra{y-x}{y}_\C.
\end{align} 
Let us take the most general state of three particles on the real line $\R$ relative to the first particle:
\begin{equation} \label{eq: generic state 3particles on a line}
    \ket{\Psi}_{\A\B\C} = \ket{0}_\A \otimes \int dx \int dy\  \psi(x,y) \ket{x}_\B \otimes \int dy  \ket{y}_\C.
\end{equation}
In the formalism of \cite{Giacomini_2019}, the reference system is not explicitly included in the state of the joint system. The equivalent state would be
\begin{equation}
    \ket{\Psi}_{\B\C} =\int dx \int dy\  \psi(x,y) \ket{x}_\B \otimes \int dy  \ket{y}_\C.
\end{equation}
When changing from reference system $\A$ to reference system $\B$, we apply the operator of \cite{Giacomini_2019}:
\begin{equation}
    \hat{S}^{\A \to \B} = \hat{\mathcal{P}}_{\A\B} e^{i/\hbar \hat{x}_\B\hat{p}_\C}.
\end{equation} 
The final state is then
\begin{align}
\hat{S}^{\A \to \B} \ket{\Psi}_{\B\C} = \int dx \int dy\ \psi(x,y) \ket{-x}_\A \otimes \ket{y-x}_\C.
\end{align}
When applying the operator in Equation \eqref{eq: app: RF change translation group} to the initial state in Equation \eqref{eq: generic state 3particles on a line}, we get the final state
\begin{align}
&U^{\A \to \B} \ket{\Psi}_{\A\B\C} \nonumber \\
=\ &\SWAP_{\A,\B}\ \circ \nonumber \\
&\left( \ket{0}_\A \int dx' dy' dx dy\ \psi(x,y) \ketbra{-x'}{x'}\ket{x}  \ketbra{y'-x'}{y''}\ket{y} \right) \nonumber \\
=\ &\ket{0}_\B \int dx \int dy\ \psi(x,y) \ket{-x}_\A \otimes \ket{y-x}_\C.
\end{align}
Hence, we see that both operators act in the same way on the most general states of particles on the real line.

\end{document}